\documentclass[pra,aps,twocolumn,groupedaddress,nofootinbib,superscriptaddress]{revtex4}
\usepackage{graphicx}
\usepackage[nointegrals]{wasysym} %
\usepackage[export]{adjustbox}
\usepackage{amsmath,amsfonts,amssymb,latexsym}
\usepackage{hhline}
\usepackage{bm}
\usepackage{verbatim}
\usepackage{enumitem}
\usepackage{mathrsfs}
\usepackage{slashed}
\usepackage{empheq}

\newcommand{\p}{\partial}

\newcommand{\lan}{\langle}
\newcommand{\ran}{\rangle}

\newcommand{\da}{{\dagger}}

\newcommand{\cmark}{\ding{51}}%
\newcommand{\xmark}{\ding{55}}%

\newcommand{\ra}{\rightarrow}

\newcommand{\uva}{{\mathbf{\hat a}}}
\newcommand{\uvb}{{\mathbf{\hat b}}}

\newcommand{\uvx}{{\mathbf{\hat x}}}

\newcommand{\bfzero}{{\mathbf{0}}}

\renewcommand{\(}{\left(}
\renewcommand{\)}{\right)}

\newcommand{\mt}{\mapsto}

\newcommand{\twp}{{2\pi}}

\newcommand{\D}{\nabla}

\newcommand\bpm            {\begin{pmatrix}}
	\newcommand\epm           {\end{pmatrix}}

\newcommand{\ms}{\medskip}
\newcommand{\bs}{\bigskip}

\def\app#1#2{%
	\mathrel{%
		\setbox0=\hbox{$#1\sim$}%
		\setbox2=\hbox{%
			\rlap{\hbox{$#1\propto$}}%
			\lower1.1\ht0\box0%
		}%
		\raise0.25\ht2\box2%
	}%
}

\newcommand{\tw}{\textwidth}

\newcommand{\vp}{\varphi}

\newcommand{\ct}{\Theta}

\newcommand{\inv}{^{-1}}

\newcommand{\ope}\odot

\usepackage{manfnt}

\newcommand{\bi}{\begin{itemize}}
	\newcommand{\ei}{\end{itemize}}

\usepackage{amsthm}

\newtheorem{theorem}{Theorem}

\theoremstyle{definition}
\newtheorem{definition}{Definition}
\theoremstyle{definition}

\newcommand\bd            {\begin{definition}}
	\newcommand\ed            {\end{definition}}
\newcommand\bt            {\begin{theorem}}
	\newcommand\et            {\end{theorem}}
\newcommand\be            {\begin{equation}}
	\newcommand\ee            {\end{equation}}
\newcommand\ba            {\begin{aligned}}
	\newcommand\ea            {\end{aligned}}
\newcommand\bea{\begin{equation}\begin{aligned}}
		\newcommand\eea{\end{aligned}\end{equation}}

\usepackage{subcaption}

\usepackage{hyperref} 
\hypersetup{final}
\hypersetup{colorlinks, citecolor=red, linkcolor=red, urlcolor=red} 

\newcommand{\sss}{\subsubsection}
\renewcommand{\ss}{\subsection}

\renewcommand{\d}{\delta}
\newcommand{\De}{\Delta}

\newcommand{\s}{\sigma}

\renewcommand{\t}{\theta}

\renewcommand{\o}{\omega}

\renewcommand{\r}{\rho}

\newcommand{\bfbe}{{\boldsymbol{\beta}}}

\newcommand{\bfl}{{\boldsymbol{\lambda}}}

\newcommand{\bfvp}{{\boldsymbol{\varphi}}}

\newcommand{\bfA}{\mathbf{A}}

\newcommand{\bfi}{\mathbf{i}}
\newcommand{\bfj}{\mathbf{j}}
\newcommand{\bfk}{\mathbf{k}}
\newcommand{\bfm}{\mathbf{m}}

\newcommand{\bfp}{\mathbf{p}}
\newcommand{\bfq}{\mathbf{q}}
\newcommand{\bfr}{\mathbf{r}}

\newcommand{\qq}{\qquad}

\newcommand{\mck}{\mathcal{K}}

\newcommand{\mco}{\mathcal{O}}

\newcommand{\mcj}{\mathcal{J}}

\newcommand{\mcl}{\mathcal{L}}

\newcommand{\mct}{\mathcal{T}}

\newcommand{\mca}{\mathcal{A}}
\newcommand{\mcp}{\mathcal{P}}

\newcommand{\mcz}{\mathcal{Z}}
\newcommand{\mcm}{\mathcal{M}}
\newcommand{\mcn}{\mathcal{N}}
\newcommand{\mcv}{\mathcal{V}}

\newcommand{\mcq}{\mathcal{Q}}

\usepackage[mathscr]{eucal} %

\usepackage{braket}

\usepackage{dcolumn}
\usepackage{pifont}
\usepackage{ulem}
\newcommand{\rcmark}{{\color{red}{\cmark}}}
\newcommand{\rxmark}{{\color{red}{\xmark}}}
\renewcommand{\rq}{{\color{red}{\mcq}}}

\begin{document}

	\title{The dipolar Bose-Hubbard model}
	
	\author{Ethan Lake}\email{elake@mit.edu}
	\affiliation{Department of Physics, Massachusetts Institute of Technology, Cambridge, MA, 02139}
	\author{Michael Hermele}

	\affiliation{Department of Physics and Center for Theory of Quantum Matter, University of Colorado, Boulder, CO, 80309, USA} 
	
	\author{T. Senthil} 
	\affiliation{Department of Physics, Massachusetts Institute of Technology, Cambridge, MA, 02139}

	\begin{abstract}
		We study a simple model of interacting bosons on a $d$-dimensional cubic lattice whose dynamics conserves both total boson number and total boson dipole moment. This model provides a simple framework in which several remarkable consequences of dipole conservation can be explored. As a function of chemical potential and hopping strength, the model can be tuned between gapped Mott insulating phases and various types of gapless condensates. The condensed phase realized at large hopping strengths, which we dub a {\it Bose-Einstein insulator}, is particularly interesting: despite having a Bose condensate, it is insulating, and despite being an insulator, it is compressible. 
	\end{abstract}
	
	\maketitle
	
	\section{Introduction} 
	
	A growing body of work has demonstrated that in systems with a conserved charge, interesting phenomena can arise if the system's dynamics conserves higher mulitpolar moments of the charge, such as dipole or quadupole moments. Systems with this type of dynamics have constrained kinematics, with the conservation laws restricting the manner in which charge is able to move. These systems have been shown to exhibit close connections with fractonic phases of matter \cite{pretko2017subdimensional,pretko2018fracton,nandkishore2019fractons,seiberg2020field,pretko2020fracton,radzihovsky2020fractons,griffin2015scalar,bidussi2021fractons,jain2021fractons,grosvenor2021space,stahl2021spontaneous}, offer ways to realize robust ergodicity breaking \cite{pai2019localization,khemani2020localization,sala2020ergodicity,rakovszky2020statistical,moudgalya2019thermalization} and anomalously slow diffusion \cite{gromov2020fracton,feldmeier2020anomalous,iaconis2021multipole,glorioso2021breakdown,grosvenor2021hydrodynamics,radzihovsky2020quantum}, and are relevant for describing experiments in systems where ultracold atoms are prepared in strongly tilted optical lattices \cite{khemani2020localization,guardado2020subdiffusion,scherg2021observing,kohlert2021experimental}.

\begin{figure}
		\includegraphics[width=.4\tw]{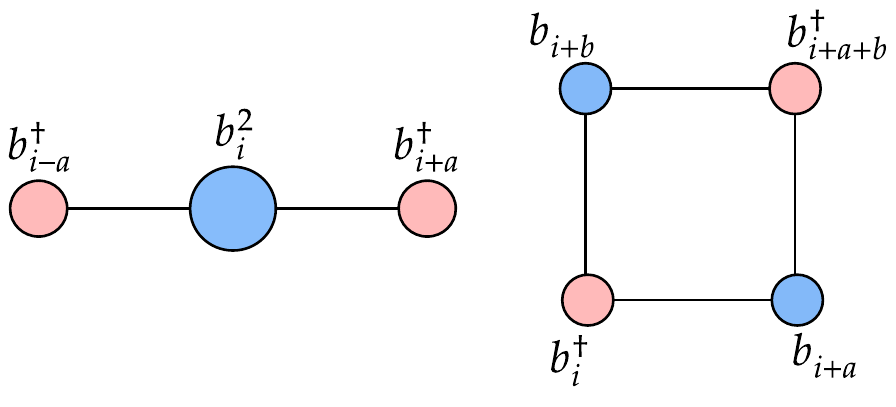} 
		\caption{\label{fig:hopping} Illustration of the simplest dipole-conserving hopping terms on the square lattice. Left: an operator that creates two dipoles with moment $\pm \uva$, separated by one lattice site in the $\uva$ direction. Right: an operator that creates two dipoles with moment $\pm \uva$, separated by one lattice site in the $\uvb\neq\uva$ direction. }
	\end{figure}

	Our aim in this work is to develop a better understanding of the physical consequences of multipolar conservation laws, and in particular to examine how such conservation laws influence the competition between kinetic energy and interactions which is at the heart of much of modern condensed matter physics. To this end, we explore a simple model that we dub the {\it dipolar Bose-Hubbard model} (DBHM), which describes interacting bosons hopping on a $d$-dimensional cubic lattice in a manner that conserves both total boson number and total boson dipole moment. The Hamiltonian of the DBHM is 
	\bea \label{ham} H_{DBHM} & = H_{hop} + H_{onsite} \\ 
	H_{hop} & = -t \sum_{i,a} b_{i-a}^\da b_{i}^2 b_{i+a}^\da - t'\sum_{i,a} \sum_{b\neq a} b_i^\da  b_{i+a} b^\da_{i+a+b} b_{i+b} \\ & \qq + h.c.\\ 
	H_{onsite} & = - \mu\sum_i n_i + \frac U2\sum_i n_i(n_i-1), \eea
	where $t,t',\mu,U$ are all positive coefficients, $n_i = b^\da_i b_i$ is the boson number operator on site $i$, and where the sums over $a,b$ run over spatial unit vectors. The hopping terms proportional to $t$ and $t'$ capture the simplest boson hopping processes compatible with dipole conservation, and are illustrated in fig. \ref{fig:hopping}. The goal in this work is to understand the competition between $H_{hop}$ and $H_{onsite}$, and by doing so to map out the quantum phase phase diagram of $H_{DBHM}$.  
	
	The conventional Bose-Hubbard model (BHM) \cite{fisher1989boson}, whose Hamiltonian is given by
	\be \label{conventional_ham} H_{BHM} =  -t_{sp} \sum_{i,a} b^\da_i b_{i+a} + h.c. + H_{onsite},\ee provides a simple model of a transition between an interaction-driven Mott insulator at small single-particle hopping $t_{sp}$, and a kinetic-energy-driven superfluid at large $t_{sp}$. This model has been extremely well-studied, and is by now textbook material \cite{sachdev2011quantum}.
	Despite the fact that the DBHM differs fundamentally from the conventional BHM only by the imposition of a single conservation law, we will see that the phase diagrams of the two models are markedly different. The large $t,t'$ phase of $H_{DBHM}$ is particularly interesting: it contains a Bose condensate and is compressible, and yet at the same time it is {\it insulating}, and displays no Meissner effect. 
	
	An outline of this paper is as follows. In section \ref{sec:mft} we discuss the mean-field phase diagram of the DBHM, which is summarized in the bottom panel of fig. \ref{fig:phase_diagram}. In section \ref{sec:cond_pheno} we explore the rather remarkable phenomenology of the condensed phase realized at large $t,t'$. Section \ref{sec:phase_transitions} is devoted to an analysis of the nature of the phase transitions identified in section \ref{sec:mft}, and in section \ref{sec:disc} we conclude with a short discussion that briefly touches on issues relevant to realizing the DBHM in experiment.

	\section{Mean-field theory and phase diagrams } \label{sec:mft}
	
	In this section we employ a simple mean-field analysis to sketch out the quantum phase diagram of the DBHM model \eqref{ham} as a function of the hopping strength and chemical potential. Because we are fixing $\mu$, we will be working in the grand canonical ensemble for boson number. We will however fix the total dipole moment, and in particular will only consider states in which it vanishes.\footnote{Focusing on such states lets us preserve the spatial symmetries of the square lattice, and states with zero dipole moment are particularly natural in the context of cold atoms, in which $H_{DBHM}$ is obtained as an effective Hamiltonian describing bosons hopping in a strongly titled optical lattice. }

\begin{figure}
		\includegraphics{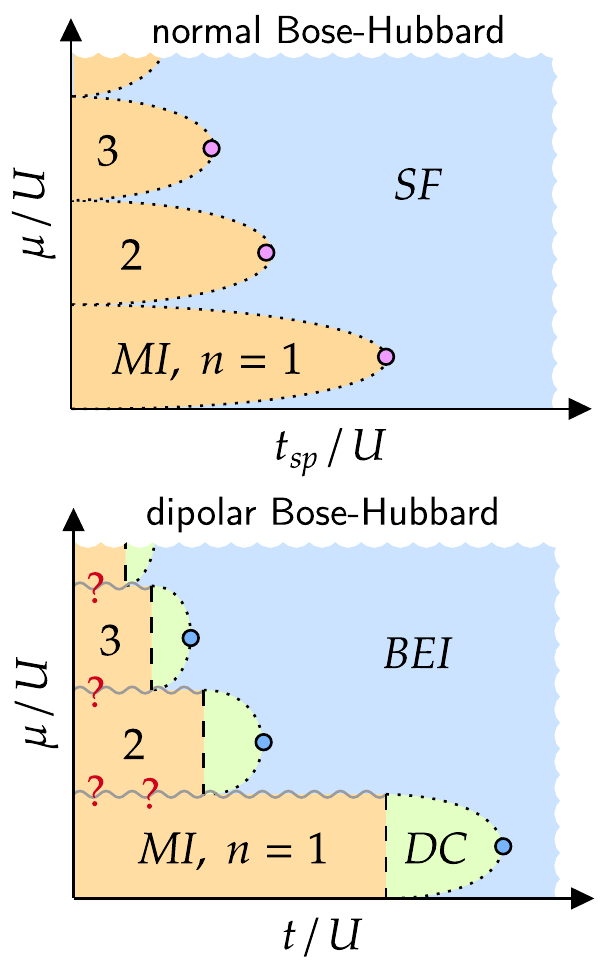}
		\caption{\label{fig:phase_diagram} Mean-field quantum phase diagram of the normal Bose-Hubbard model (top) and the dipolar Bose-Hubbard model (bottom). The orange regions are Mott insulators, with the integer labels denoting the average density of bosons on each site. The green regions are dipole condensates, where dipoles (but not single bosons) have condensed; the average density of these phases is the same as that of their parent Mott insulators. The blue region in the top plot is a conventional superfluid, while the blue region in the bottom plot is a `Bose-Einstein Insulator' (BEI) --- a gapless condensate with dynamical exponent $z=2$ discussed in sec. \ref{sec:cond_pheno}. 
			Dashed lines denote transitions which are either first-order or are described by $(d+1)$-dimensional XY models, and dotted lines denote transitions described by the $d$-dimensional dilute Bose gas. The purple circles in the top plot denote multicritical points described by the $(d+1)$-dimensional XY model, while the blue circles in the bottom point denote the multicritical points discussed in sec. \ref{sec:phase_transitions}. 
			{Red question marks in denote various types of crystalline states beyond the present mean-field framework. At large $t/U\gtrsim 1$ and small $n$, the BEI phase is unstable to the clumped phase (not shown) discussed in the main text. }
		}
	\end{figure}

	Before starting, we briefly recapitulate the physics of the conventional BHM \cite{fisher1989boson}. The phase diagram of this model is reviewed in the top panel of fig. \ref{fig:phase_diagram}. When the single-particle hopping $t_{sp}$ vanishes, the system forms a gapped Mott insulating state, with the average boson number $n$ at each site quantized to be an integer determined by the ratio $\mu/U$. 
	Increasing $t_{sp}$ lowers the gap to doped particles via virtual processes in which single particles delocalize around the lattice. When the decrease in energy brought about by these hopping processes brings the gap to zero, the doped particles condense to form a superfluid. 
	
	The transition between the Mott insulators and the superfluid takes place along a series of dome-shaped critical lines. The transition generically occurs when particles (or holes, depending on the value of $\mu$) are gradually doped into the parent Mott insulator, and in this case the critical point is described by a dilute Bose gas of particles (or holes), with a dynamical exponent of $z=2$, and with the average density changing smoothly across the transition.
	This story is modified at the ``tips'' of the Mott insulating regions (purple circles in the top panel of fig. \ref{fig:phase_diagram}). At these multicritical points the energy gaps to doped particles and doped holes simultaneously vanish, thereby producing an effective particle-hole symmetry. In this case the average density is unchanged across the transition, which has $z=1$ and which is described by the critical point of the $(d+1)$-dimensional XY model. 
	
	With this review out of the way, let us now turn our attention to the DBHM.
	The $t=t'=0$ limit of the DBHM Hamiltonian \eqref{ham} is the same as the $t_{sp}=0$ of limit of the regular BHM, and consequently in this limit we obtain a series of Mott insulators with fixed integral average particle number per site. 
	
	In the opposite limit of large $t/U, t'/U$, the DBHM develops an instability towards states in which the bosons form isolated clumps with large boson number. This is because when acting on a region with local density $n$, the hopping operators $-b_{i-a}^\da b_i^2 b_{i+a}^\da, \,\, -b^\da_i b_{i+a} b_{i+a+b}^\da b_{i+b}$ have eigenvalues that scale with $n$ as $-n^2$, which has the same scaling as the Hubbard repulsion $\sim U n^2$. When $t,t'$ are sufficiently larger than $U$, the energy can thus always be lowered by making the local boson density as large as possible, with the instability arrested only by higher-order interactions like $n^4$ (note that this instability would not arise in analogous dipole-conserving {\it spin} models). The physics of this clump formation (which is similar to the fractonic microemulsions of Ref. \cite{prem2018emergent}) will be discussed elsewhere \cite{lake20221dmodel}, and in the remainder of this paper we will focus on the intermediate $t/U,t'/U$ regime, where as we will see several exotic stable phases of matter exist.
	
	Consider then what happens to the Mott insulators as the boson hopping terms are turned on. 
	The kinematic constraint imposed by the dipolar conservation law prevents a superfluid from forming in the way that it does in the conventional BHM. Indeed, consider what happens when one dopes particles into a given Mott insulator. Due to dipole conservation, an isolated doped particle is {\it completely immobile}, and cannot lower its energy through any processes which do not create additional excitations. 
	
	\begin{figure}
		\includegraphics[width=.45\tw]{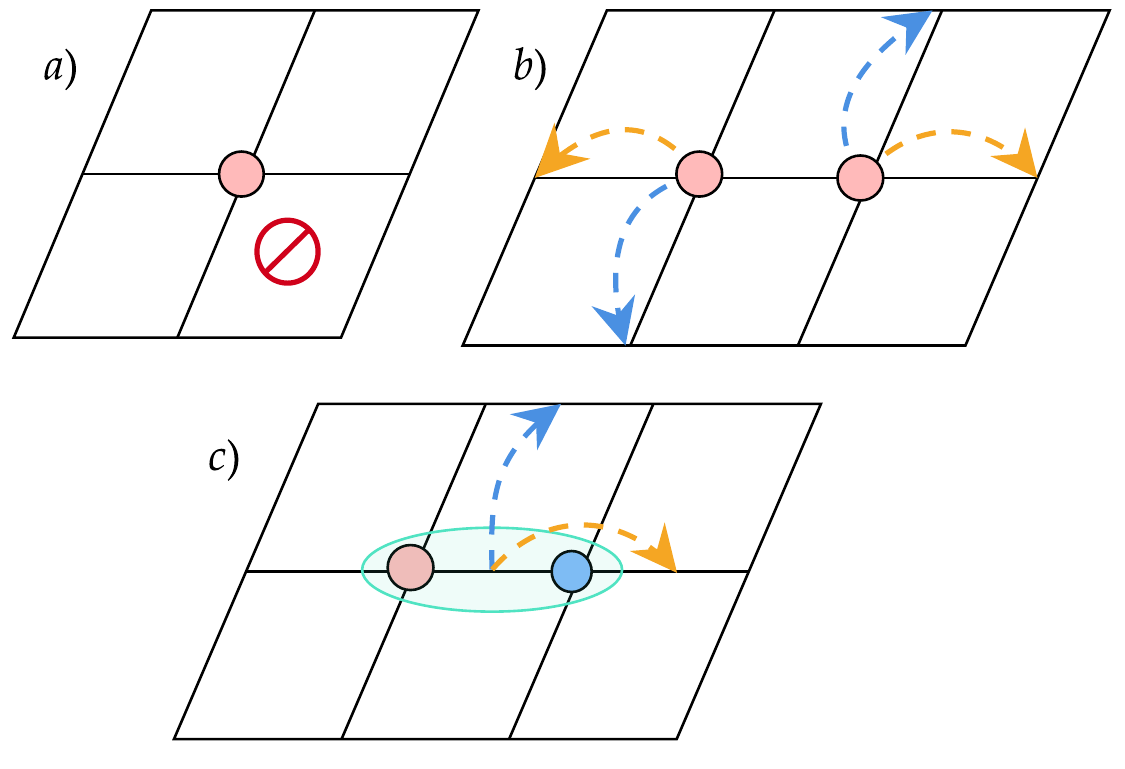} 
		\caption{\label{fig:kin_constraints} An illustration of the kinetic constraints imposed by dipole conservation. $a)$ A single isolated boson (or hole) doped into a given Mott insulator is completely immobile. $b)$ two bosons located near one another can undergo collective motion which preserves their center of mass, where each boson hops in the $\pm \uva$ direction. $c)$ A dipolar bound state of a boson and a hole is free to move in any direction. }
	\end{figure}

	Now consider a pair of nearby doped particles. These particles are able to move in a restricted sense, since they can ``push'' off of one another and move in opposite directions --- see panels $a$ and $b$ of fig. \ref{fig:kin_constraints} for an illustration. However, once the distance between the particles becomes significantly larger than the range of the hopping terms in the Hamiltonian, their motion will again be frozen out. This means that doped particles will only be able to fully delocalize throughout the lattice when their average density is of order 1.
	
	This suggests that a transition out of the Mott insulator which proceeds directly by condensing single bosons will be first order, as the kinematic constraint means that it is impossible for the bosons to delocalize at arbitrarily small doping levels --- the density must therefore jump discontinuously across any direct transition where single bosons condense. In line with these expectations, a naive single-boson mean field treatment of \eqref{ham} (performed by writing $b_i = (b_i - \psi) + \psi$ and working to lowest order in $(b_i-\psi)$) indeed generically yields a first-order transition. 
	
	As we discuss in Sec. \ref{sec:cond_pheno}, the resulting condensed phase is actually very interesting. Before explaining why, we will take a moment to analyze a more natural two-step condensation process, whereby boson condensation is mediated by the condensation of charge-neutral dipolar `exciton' bound states.

	\ss{Dipole condensation} 
	
	The most natural way of transitioning out the Mott insulator can be understood by noting that the hopping terms in \eqref{ham} are simply conventional kinetic terms for the operators $b_i b_{i+a}^\da$, which create dipolar bound states with dipole moment along $\uva$. This means that while isolated doped particles are localized by the kinematic constraint, dipolar bound states can move freely in all directions (see panel $c$ of fig. \ref{fig:kin_constraints}). Since dipole motion is the easiest way for the system to lower its kinetic energy, we expect dipole condensation to occur before single-particle condensation does. Note that since the dipoles are charge neutral, the resulting dipole condensate possess the same average density as its parent Mott insulator.\footnote{Here, as well as in all of what follows, we will restrict our attention to positive values for $\mu$, where the $t=0$ Mott insulators have a nonzero average density. For $\mu<0$ by contrast the $t=0$ state has no particles at all, and it is therefore impossible to create dipoles on top of the ground state. In this case, a direct first-order transition across which single bosons condense seems to be the most natural outcome, but we leave a more detailed investigation to future work. } 
	
	We may analyze this transition within a simple mean-field framework by decoupling the quartic terms in $H_{hop}$. To do this, we write $H_{hop}$ as
	\be H_{hop} = -\sum_{i,a} (d^a_i)^\da \mca^a_{ij} d^a_j,\ee 
	where we have defined the operators 
	\be d^a_j \equiv b_{j+a}^\da b_j,\ee 
	which create boson configurations with dipole moment along the $\uva$ direction,
	as well as the matrix 
	\be [\mca^a]_{ij} \equiv \sum_b \(t\d_{a,b} + t' (1-\d_{a,b}) \)(\d_{i,j+b} + \d_{i,j-b}).\ee
	Decoupling $H_{hop}$ by introducing a set of dipole fields $D_i^a$, we obtain an imaginary-time coherent-state path integral with the action
	\bea S &= \int d\tau \Big( \sum_{i} b^\da_i\p_\tau b - \sum_{i,a} \( (d^a_i)^\da D^a_i + (D^a_i)^\da d^a_i\) \\ 
	& \qq + \sum_{i,j,a} (D^a_i)^\da [\mca^a]\inv_{ij} D_j^a + H_{onsite}\Big),\eea
	with the dipole fields satisfying  
	\be \lan D^a_i\ran = \sum_j \mca^a_{ij} \lan d^a_j\ran  = \sum_b \lan d^a_{i-b} + d^a_{i+b}\ran(t\d_{a,b} + t'(1-\d_{a,b})).\ee 
	
	We now play the usual game of integrating out the $b$ fields to generate an effective continuum action for the $D^a$ dipole variables. On general grounds the most relevant pieces of such an effective action compatible with dipole conservation and spatial symmetries may be written as 
	\bea \label{spheno} S_{eff} & = \int d\tau \, d^dx\, \Big(  w\sum_a  |\p_\tau D^a|^2  + r\sum_a |D^a|^2 \\ & \qq  + \sum_{a,b} \frac{K_D^{ab}}2 |\D_aD^b|^2 + \sum_{a,b} \frac{ u^{ab}}2 |D^a|^2 |D^b|^2\Big). \eea 
	Here the absence of a linear time derivative term follows under the spatial reflection symmetry which sends $D^a_i$ to $(D^a_i)^\da$: a nonzero density of $D^a$ dipoles would break reflection symmetries that send $\uva$ to $-\uva$, and hence in the presence of such symmetries the dipoles must all be at zero density. 
	
	At the mean field level, the transition occurs when $r=0$. Expressions for the coefficients in $S_{eff}$ in terms of the microscopic parameters in \eqref{ham} can be calculated using standard methods (see appendix \ref{app:effective_action}); in particular for $r$ one finds 
	\bea \label{r} 
	r & = \frac1{2t + 2(d-1)t'}-\frac{2n(n+1)}{U},
	\eea 
	so that e.g. when $t = t'$, the transition occurs at $t_* = U / (4dn(n+1))$. Note that when $n$ is large, this value allows for a clear parametric separation between the onset of the dipole condensate and the instability caused by the clumping phenomenon mentioned earlier, which occurs when $t \gtrsim U$. The extent of this separation at small $n$ is a question requiring a more detailed numerical treatment \cite{lake20221dmodel}; in what follows we will simply assume that a nonzero separation exists (which can always be arranged by working at large $n$). 
	
	The first term in \eqref{r} arises from the zero-momentum contribution to $(D^a)^\da [\mca^a]\inv D^a$, while the second piece arises from performing perturbation theory in the $b^\da_i b_{i+a} D^a_i +h.c.$ term. The $U$ in the denominator of this piece comes from the energy cost of creating a particle-hole pair on top of the Mott insulating ground state. Since this energy is independent of $\mu$ to second order in perturbation theory, the shape of the mean-field phase boundary separating the Mott insulator from the dipole-condensed state is {\it independent} of $\mu$ to leading order (see fig. \ref{fig:phase_diagram}). In particular, at fixed small values of $t,t'$, changing $\mu$ will simply induce a direct first-order transition between different Mott insulators, at least at the level of the mean field analysis considered here. 
	{Going beyond the present mean field approximation, the regions in between distinct Mott insulators will likely host various intermediate density states that spontaneously break lattice symmetries.}
	\footnote{We thank David Huse for this remark.} 
	{We leave a detailed investigation of these states to future work.}
	
	Now let us now address the nature of the dipole condensate that forms when $r<0$.
	Let us write $u^{ab} = u_d\d^{ab} + u_o(1-\d^{ab})$, with $u_d>0$ assumed to be positive. 
	If $u_o>u_d$, the system prefers to break the discrete lattice rotational symmetry and condense only a single species of dipole, with the condensed phase possessing a single gapless mode. 
	If $u_o<u_d$ on the other hand, the system prefers to condense dipoles of all orientations with equal magnitudes. This happens though a continuous transition if $u_o > -u_d/(d-1)$; otherwise the potential for the dipole fields is unbounded from below to quartic order, and the transition is likely to be rendered first order (which is in fact what happens within a mean-field analysis for the particular Hamiltonian in \eqref{ham}; see appendix \ref{app:effective_action} for details). 
	
	Note that even in the case where all species of dipoles condense, the condensate generically spontaneously breaks both lattice reflections (unless ${\rm Im}[\lan D^a\ran]=0$ for all $a$) and lattice rotations (unless $\lan D^a\ran$ is independent of $a$). Translation symmetry is unbroken in the condensate however, since the operators which condense are the zero-momentum components of $D^a$.

	\ss{Single boson condensation} 
	
	After dipoles have condensed, there is no longer any kinematic obstruction to condensing single bosons, since the presence of the dipole condensate eases the kinematic constraint --- roughly speaking, single bosons are now free to move by absorbing dipoles from the condensate. 
	In the case where all species of dipole condense with equal magnitudes, we may write 
	\be D^a = \sqrt{\r_D} e^{i\vp_D^a},\ee 
	with $\r_D$ a nonzero constant determining the dipole condensate fraction.	
	This substitution yields the effective Hamiltonian 
	\bea \label{dcondensed_ham} H & = \sum_{i,a} \sqrt{\r_D} \((d^a_i)^\da  e^{i\vp_D^a} + d^a_ie^{-i\vp_D^a}\) + H_{onsite} \\ 
	&+ \frac{K_L}2 (\D\cdot \bfvp_D)^2 + \frac{K_T}2(\D\times\bfvp_D)^2 + \frac{K_A}2 \sum_a (\D_a \vp^a_D)^2, \eea 
	where $K_L,K_T,K_A$ set the stiffness for the phase modes of the dipole condensate, with the anisotropic term $K_A$ allowed by the cubic lattice symmetry.
	The first term proportional to $\sqrt{\r_D}$ provides an effective single-boson hopping term (recall $d^a_i = b_{i+a}^\da b_i$), and since 
	$\r_D$ increases as one goes further into the dipole condensed phase, eventually --- at least in 2 and 3 dimensions --- one triggers a transition at which single bosons condense. In 1d the situation is slightly different, as the effects of vortices in the boson phase need to be taken into account. A detailed analysis of these effects will be given in \cite{lake20221dmodel}, and in the following we will simply restrict to $d>1$, where vortices can effectively be ignored for the present purposes. 
	
	The location of the phase boundary where single bosons condense can be determined by performing single-particle mean field theory on \eqref{dcondensed_ham} in the standard way \cite{fisher1989boson};\footnote{In the presence of a dipole condensate, the ground state to perturb about is not given by the usual Mott-insulating ground state, although it does have the same average density. However, in the limit where the condensate fraction of the dipoles is small, treating the ground state as the Mott insulating one will still give accurate results for the phase boundary.} this gives rise to the domed parts of the phase diagram in the bottom panel of fig. \ref{fig:phase_diagram}. We note in passing that this series of transitions --- where an intermediate dipolar condensed phase separates the Mott insulating and single boson condensed phases, and provides a way for single bosons to move --- is conceptually quite similar to the theory of 2d dislocation-mediated quantum melting put forward in \cite{kumar2019symmetry,pretko2018symmetry,radzihovsky2020quantum,zhai2021fractonic}.
	
	When the interaction matrix $u^{ab}$ is such that only one species $D^a$ of dipole condenses, single boson hopping is only generated along a single direction, and the condensed phase possesses a quasi-1d character. The consequences that this quasi-1d behavior has for the nature of the condensed phase and the character of the condensation transition are discussed in appendix \ref{app:single_dipole_cond}. For simplicity, in the rest of the main text we will specialize to the case where all species of dipoles condense. 
	
	Recall that in the dipole condensed phase, the symmetries of lattice rotations and reflections are generically spontaneously broken, while translation symmetry is preserved. In the single-boson condensed phase however, we have (assuming that the condensate is not destroyed by fluctuations; the criteria for when this happens will be discussed in the next section) 
	\be \label{bcond} \langle b_i\rangle = |\lan b \ran| e^{i\alpha + i\bfbe\cdot \bfi},\ee 
	where $\lan e^{i\bfvp_D} \ran \sim e^{i\bfbe}$ and where $|\lan b_i\ran|$ is independent of the lattice site $i$. 
	In particular, as long as $\bfbe$ is nonzero, translation symmetry is spontaneously broken in the condensate. However, a subgroup mixing global charge conservation and translation symmetries is preserved, as \eqref{bcond} is left invariant under the transformation 
	\be b_i \mt b_{i+\uva} e^{-i\bfbe\cdot\uva}.\ee 
	Thus the single-particle condensed phase realizes a type of spiral ordering, with intertwined patterns of phase and translational ordering. Note however that spirals with different pitch are in fact degenerate in energy, as they are related by shifts of $\bfbe$.

	\section{Phenomenology of the single-particle condensed phase: the Bose-Einstein insulator}\label{sec:cond_pheno}
	
	\ss{IR field theory and symmetry breaking} 
	
	In this section, we explore the phenomenology of the phase in which single particles have condensed (still restricting our attention to $d>1$). 
	In this phase, the resulting IR theory may be written in terms of the phase mode $\phi$ appearing in $b \sim e^{i\phi}$ as\footnote{The term $K_2(\D^2\phi)^2/2$ is also allowed by symmetry, but as we will see shortly $K_2$ merely leads to a renormalization of $K_L$.}
	\bea \label{condensed_mcl} \mcl & = \frac \kappa2(\p_\tau \phi)^2 + \frac{w}2(\p_\tau \bfvp_D)^2 + \frac{K_\r}2(\D\phi - \bfvp_D)^2 \\ 
	& + \frac{K_L}2(\D \cdot \bfvp_D)^2 + \frac{K_T}2(\D\times\bfvp_D)^2 + \frac{K_A}2 \sum_a (\D_a\vp_D^a)^2, \eea 
	where $K_\r$ is a stiffness parameter proportional to $\sqrt{\r_D}$. 
	
	We see from \eqref{condensed_mcl} that the single-particle condensate ``Higgses'' the dipole Goldstone $\bfvp_D$: we may shift $\bfvp_D \mt \bfvp_D+ \D\phi$, with the term proportional to $K_\r$ then effectively gapping out the dipolar phase field, allowing us to set $\bfvp_D=0$. That $\bfvp_D$ disappears from the IR theory is of course completely physical: in the presence of a single-particle condensate, the phase of $\lan d_i^a\ran = b^\da_{i}b_{i+a}^\da\ran$ is no longer an independent variable, and is determined by the phase of $\lan b_i\ran$ --- thus only $\phi$ (and not $\bfvp_D$) should be a low-energy degree of freedom in the IR theory. 
	
	With $\bfvp_D$ out of the way, we may thus write (dropping $K_T(\D\times \D \phi)^2/2$, which vanishes away from vortices) 
	\bea \label{z2action} \mcl & = \frac \kappa2(\p_\tau \phi)^2 + \frac{K_L}2(\D^2 \phi)^2 + \frac{K_A}2 \sum_a (\D_a^2\phi)^2. \eea 
	This is the Lagrangian of an anisotropic quantum Lifshitz model (QLM), which has also appeared in the analysis of the fractonic ``superfluids'' of refs. \cite{yuan2020fractonic,chen2021fractonic}.\footnote{The one-dimensional version of this model has also recently studied in ref. \cite{gorantla2022global}.}
	In most applications the QLM is realized only at a critical point \cite{fradkin2004bipartite,zhang2016superfluid,lake2021re,ma2018higher}, arising when the coefficient of a single gradient term $(\D \phi)^2$ is tuned through zero. By contrast, the QLM written down above describes an entire phase of matter, made possible by the dipolar symmetry which forbids the aforementioned single-gradient term.\footnote{In the normal QLM, the critical point in e.g. $d=2$ has the possibility of being destabilized by marginal terms such as $(\D \phi)^4$ \cite{vishwanath2004quantum,fradkin2004bipartite}. In the present setting however such terms are forbidden by dipole conservation, and these issues do not arise.}

	Let us now examine the pattern of symmetry breaking that occurs in the condensed phase.
	The equal-time $T=0$ boson two-point function, which in the IR is determined by the two-point function of $e^{i\phi}$, is 
	\bea \label{vertex_correlator} \lan e^{i\phi(\bfr)} e^{-i\phi(\bfzero)} \ran = \exp\( -\int \frac{d\o\, d^dk}{(\twp)^{d+1}}\frac{1-\cos(\bfk\cdot\bfr)}{\kappa\o^2+K_L k^4}\),\eea
	where we have momentarily set $K_A=0$ for simplicity.
	By looking at the small $k$ behavior of the integral as $r\ra\infty$, we see that $e^{i\phi}$ has QLRO in $d=2$, and LRO in $d=3$ \cite{yuan2020fractonic,stahl2021spontaneous}. 
	
	At finite temperatures, the integral over $\o$ is replaced with a Matsubara sum, and $e^{i\phi}$ is seen to have short-range correlations in {\it all} $d\leq 3$. In $d<3$, vortices (textures around which $\phi$ winds by $\twp$, with a core at which $|\psi|\ra0$) proliferate at any nonzero $T$, which happens due to the fact that because of the structure of the kinetic term, a single isolated vortex does {\it not} cost a thermodynamically large amount of gradient energy.\footnote{Thus in $d=2$, $T_{BKT} = 0$. This is also true in the normal QLM \cite{ghaemi2005finite}, but for more subtle reasons.} Thus at any $T>0$ the field $\phi$ ceases to be well defined and the IR theory is trivial, with all operators exhibiting short-ranged correlation functions. 
	In $d=3$ by contrast, although $e^{i\phi}$ is disordered at any $T>0$, vortex lines do not immediately proliferate, since they are extended objects possessing a nonzero core energy. In this case $\phi$ remains well-defined at small $T$ and the IR theory remains nontrivial, with many operators possessing long-range correlation functions (such as $e^{i\D_a \phi}, \,e^{i\p_\tau \phi}$). At large enough $T$, the vortices lose their line tension, and the theory passes into the trivial disordered phase by way of an inverted 3d XY transition.

	\begin{table}
		\renewcommand{\arraystretch}{1.4}
		\begin{tabular}{@{}llll@{}}  
			
			dip. cond.\,\, & $d=1$\,\, & $d=2$ \,\, & $d=3$   \\ \hline 
			
			$U(1)_D$ & $\mcq$, \rcmark  & \xmark, $\rq$ & \xmark , \rxmark\\ 	
			
		\end{tabular}
		
		\ms 
		
		\begin{tabular}{@{}llll@{}}  
			
			boson cond.\,\, & $d=1$\,\, & $d=2$ \,\, & $d=3$   \\ \hline 
			
			$U(1)_D$ &  n/a  & \xmark, \rcmark & \xmark , \rxmark\\ 	
			$U(1)_C$ & n/a & $\mcq$, \rcmark & \xmark, \rcmark \\
			
		\end{tabular}
		
		\caption{\label{tab:ssb} The symmetry breaking patterns in the dipole condensed phase (top) and single-particle condensed phase (bottom). Here $U(1)_D$ indicates the dipole symmetry, and $U(1)_C$ indicates global charge conservation. A \cmark\, indicates that the symmetry is preserved while a \xmark\, indicates that SSB occurs, with $\mcq$ standing for a symmetry-preserving phase with QLRO. The black symbols give the symmetry breaking pattern at $T=0$, with the red symbols corresponding to $T>0$. $U(1)_C$ is always preserved in the dipole condensed phase, and both symmetries are always preserved in the Mott insulating phase. We write `n/a' in the second table to signify that understanding the 1d theory in this case requires dealing with the effects of vortices \cite{lake20221dmodel}, which goes beyond the present analysis.}
	\end{table}
	
	We may also consider correlation functions of dipole operators, which in the IR map to correlators of $e^{i\D _a\phi}$. The correlators are calculated as in \eqref{vertex_correlator}, except with an extra factor of $1-\cos(\bfk\cdot\uva)$ appearing in the integrand. This extra factor means that the integral is never IR-divergent, so that at $T=0$, $e^{i\D_a \phi}$ {\it always} has long-range order in the boson condensed phase, even in those cases for which $e^{i\phi}$ is disordered. 
	The patterns of symmetry breaking that occur throughout the phase diagram are summarized in table \ref{tab:ssb}.

	\ss{Electromagnetic response} 
	
	Let us now address whether or not the boson condensed phase is as a superfluid.
	If we were to define a superfluid as a state in which global particle number conservation is either spontaneously broken or has QLRO, the condensed phase would count as a superfluid for $d = 2,3$ (provided that $T=0$). This however is not the correct definition of a superfluid: a more precise definition (albeit one that is frequently used interchangeably with the above statement about symmetry breaking) is a system with a nonzero superfluid weight, viz. a system which exhibits the Meissner effect when coupled to a background electromagnetic field. In this sense, the boson condensed phase is {\it not} a superfluid, in any dimension. Indeed, the superfluid weight of the boson condensed phase {\it vanishes}, and it does not display any Meissner effect. 
	Most strikingly, the condensed phase is in fact completely insulating, at any temperature.\footnote{That systems with dipole conservation must have zero DC conductivity $\s_{DC}$ (even if the dipole symmetry is spontaneously broken) is essentially due to the fact that dipole conservation prevents motion of the center of mass of the charge carriers. This argument can be made more rigorous by an analysis similar to the one employed in the discussion of Bloch's theorem in \cite{else2021critical}.} 
	
	To understand these statements, we simply observe that a background field $A_\mu$ for the global boson number symmetry couples to the phase action as 
	\bea \label{gaugedz2action} \mcl & = \frac \kappa2(\p_\tau \phi-A_0)^2 + \frac{K_L}2 (\D^2\phi-\D\cdot \bfA)^2 \\ & \qq + \frac{K_A}2\sum_a (\D_a^2 \phi -\D_aA_a)^2, \eea 
	which can be derived by sending $\p_\mu\phi \ra \p_\mu \phi - A_\mu$ and $\bfvp_D \ra \bfvp_D-\bfA$ in \eqref{condensed_mcl} (this procedure also produces the unimportant term $K_T(\D \times \bfA)^2/2$, which simply renormalizes the electric charge).
	
	We see from \eqref{gaugedz2action} that no mass is generated for the vector field $\bfA$, since only spatial derivatives of $\bfA$ appear in the above Lagrangian --- there thus is no Meissner effect, and both the DC conductivity and superfluid weight vanish, {which follows simply from the absence of the $(\D \phi)^2$ term in \eqref{z2action}}. Note that this occurs despite the fact that particle number conservation is spontaneously broken (at least in $d=3$). We have thus realized a rather remarkable scenario wherein even though bosons are condensed, the system is insulating, and incapable of transporting charge. Furthermore, as the average density changes continuously in the condensed phase, this system provides an example of a very unusual phase of matter: a translationally-invariant compressible insulator!  For these reasons, we will refer to the single-particle condensed phase as a {\it Bose-Einstein insulator} (BEI). 
	
	We note as an aside that compressible translation-invariant systems (without a microscopic dipole conservation symmetry) with nonzero resistivity at $T=0$ were recently studied in \cite{else2021critical}, where  they were made possible by a phenomenon the authors dubbed ``critical drag''. As explained in Ref. \onlinecite{else2021critical}, critical drag is operative in the model \eqref{z2action} at $T=0$, which implies that $\s_{DC}$ must vanish at zero temperature. The fact that in the present setting we actually have $\s_{DC}=0$ for {\it all} $T$ is a consequence of the assumed microscopic dipole symmetry.   
	
	Furthermore, it is not just the DC conductivity of the BEI that vanishes. Indeed, the electromagnetic response kernel derived from \eqref{gaugedz2action} is 
	\bea\label{emresponse} - & \frac{\d^2 \ln \mcz[\bfA]}{\d A^a_{\o,\bfq} \d A^b_{-\o,-\bfq}}\Big|_{\bfA = 0}  = q_aq_b(K_L + K_A \d_{ab}) \\ 
	& \qq - q_aq_b \frac{(K_Lq^2 + K_Aq_a^2)(K_Lq^2 + K_Aq_b^2)}{\kappa \o^2 + K_L q^4 + K_A \sum_c q_c^4}, \eea
	whose transverse part vanishes in the zero-momentum limit, in accordance with the vanishing of the superfluid weight.  
	The kernel in \eqref{emresponse} implies that the conductivity $\s(\o,\bfq)$ vanishes at zero momentum for {\it all} frequencies, viz. 
	\be \label{zerosig} \s(\o,\bfzero) = 0.\ee 
	This fact is a simple consequence of dipole symmetry, and holds even if the $U(1)$ particle number symmetry is spontaneously broken. Indeed, if $\rho(x)$ is the charge density and $J^i(x)$ is the charge current, then 
	\be [H,\int d^dx\, x^i\rho(x)] =-i \int d^dx\, x^i \D_j J^j(x) = i\int d^dx\, J^i(x).\ee 
	Since this vanishes by dipole conservation, we have that $J^i=0$ in any state with uniform current density, thus implying the vanishing of the conductivity in \eqref{zerosig} {(note however that the conductivity associated with a rank-2 gauge field that couples linearly to the {\it dipole} current will be nonzero)}.
	
	The finite-momentum conductivity by contrast is generically nonvanishing (although the $\o=0$ response vanishes at all $\bfq$ unless $K_A\neq0$), and even contains a $1/(\omega + i0)$ Drude-type pole. Thus the BEI can behave like a superconductor at short distances, but is nevertheless insulating at the longest length scales. 
	
	Since the real part of the conductivity extracted from \eqref{emresponse} goes as $q^2$ at small frequencies, and since the compressibility of the BEI is nonzero, the charge dynamics in the BEI is subdiffusive, with a diffusion ``constant'' going as $q^2$. The transition between this subdiffusive behavior and the fully gapped charge response of the Mott insulating and dipole condensed phases could potentially be used as a way to identify the BEI in experiment, {with subdiffusion being probed by examining the evolution of the density following a quench, as in \cite{scherg2021observing,guardado2020subdiffusion}}.

	\section{Nature of the phase transitions} \label{sec:phase_transitions}
	
	We turn now to analyzing the nature of the phase transitions identified within the above mean-field framework. When $d>1$ we will continue to assume that all species of dipoles condense, with the case where only a single dipole condenses being treated in appendix \ref{app:single_dipole_cond}.
	
	\ss{Dipole condensation transitions} 
	
	We first address the simpler case of the transitions that occur when dipoles condense out of a Mott insulator (dashed lines in the bottom panel of fig. \ref{fig:phase_diagram}). {For simplicity we will only consider the case where the dipole hopping is isotropic in space, so that $K_D^{ab}$ in \eqref{spheno} is independent of $a,b$.}

	For $d=1$, the transition into the dipole condensed phase is simply that of the $2$-dimensional classical XY model. 
	When $d=2$, {one possible critical point is given by two copies of the $3$-dimensional classical XY transition (see also \cite{calabrese2004critical})}. The most relevant couplings couple the energy operators on each copy, and are (barely) irrelevant \cite{poland2019conformal}: hence a transition described by two decoupled $3$-dimensional classical XY models can occur. In $d=3$ the quartic couplings between the dipole fields are marginally irrelevant if positive (hence yielding a stable decoupled fixed point with mean-field exponents), while if they are sufficiently negative they can be made marginally relevant, likely producing an instability towards a first-order transition. 
	
	\ss{Single particle condensation} 
	
	More interesting transitions occur when single bosons condense on top of a background dipole condensate (dotted lines in the bottom panel of fig. \ref{fig:phase_diagram}). 
	The effective field theory describing the transition has the Lagrangian 
	\bea \label{singlepl} & \mcl  = s\psi^\da \p_\tau \psi + p |\p_\tau\psi|^2 - \mu |\psi|^2 + \frac u2|\psi|^4 \\ & \qq + \r_D |(\D - i\bfvp_D)\psi|^2 + \mcl_0[\bfvp_D],\eea 
	where $\psi$ is a complex field and $\mcl_0[\bfvp_D]$ contains the Gaussian terms for $\bfvp_D$. 
	Note that $\bfvp_D$ enters the kinetic term for $\psi$ in the way that an electromagnetic gauge field would, with the structure of the derivative coupling $|(\nabla - i\bfvp_D)\psi|^2$ fixed by dipole symmetry. Unlike a gauge field however the kinetic term for $\bfvp_D$ is not invariant under shifts of $\bfvp_D$ by total derivatives, and there is no corresponding electric potential appearing in the $\psi^\da\p_\tau \psi$ term. 
	
	The nature of the critical point where $\psi$ condenses depends on the spatial dimension $d$, as well as whether or not the transition is generic (occurring at varying density; $s\neq0$) or multicritical (occurring at fixed density; $s=0$). Since the analysis in the case of $d=1$ requires understanding the effects of vortices \cite{lake20221dmodel}, we will restrict to $d=2,3$ in what follows. 
	
	\sss{$s\neq0$}

	Consider first the generic transition with $s\neq0$, where the $p|\p_\tau\psi|^2$ term is irrelevant and may be dropped. In this case only particles or holes (but not both) are doped into the dipole-condensed phase, and the density changes continuously across the transition. In the absence of the coupling to $\bfvp_D$, these transitions would be described by the $d$-dimensional dilute Bose gas. 
	
	Consider first $d=3$. Under $z=2$ scaling, the coupling between $\psi$ and $\bfvp_D$ is irrelevant; consequently the critical point is simply that of the dilute Bose gas. 
	
	In $d=2$, $u$ is marginal. Taking $K_L = K_T\equiv K,K_A=0$ for simplicity, the flow to leading order in $1/K$ and $u$ is 
	\be \frac{du}{dt} = -\frac{A}{K^2} - Bu^2,\ee 
	where $A$ and $B$ are positive constants. Since $1/K$ is always marginal (the self energy of $\bfvp_D$ is trivial on account of there being no production of virtual $\psi$ particles), the first term means that $u$ is always eventually driven negative, implying that the transition is likely to generically be rendered first-order. 
	
	
	On the condensed side of the transition, the usual mean-field Bogoliubov treatment gives a single mode with dispersion (setting $s=1$ and taking $K_D^{ab} = K_D$ independent of $a,b$ for simplicity)
	\be\label{bogol} \o(\bfk) = \frac{\o_B(\bfk)}{\sqrt{1+2\r_D\lan |\psi|^2\ran/(K_Dk^2)}},\ee 
	where $\o_B(\bfk) =\sqrt{k^2\r_D}\sqrt{k^2\r_D+2\mu}$ is the familiar Bogoliubov dispersion of the condensate in the absence of the coupling to $\bfvp_D$, and where the square root factor on the RHS of \eqref{bogol} comes from hybridization with $\bfvp_D$. By examining the small $k$ limit of the above expression, we see that the coupling to $\bfvp_D$ correctly produces the $z=2$ dispersion of the BEI, instead of the $z=1$ of conventional superfluids. 
	
	\sss{$s=0$}
	
	We now examine the nature of the multicritical points where $\mu$ is tuned to ensure particle-hole symmetry about the given Mott insulating ground state, so that the condensation transition occurs at fixed density. In the absence of the $\bfvp_D$ field, these transitions would be described by the critical point of the $(d+1)$-dimensional classical XY model. 
	
	For $d=3$, the coupling to $\bfvp_D$ renders the transition first order, via essentially the same mechanism as in 3d scalar QED \cite{coleman1973radiative,halperin1974first,halperin1974analogy}. 
	
	For $d=2$, the Lagrangian \eqref{singlepl} with $s=0$ is in fact exactly equivalent to the field theory describing the nematic to smectic-A transition, upon identifying imaginary time with the spatial direction normal to the smectic planes and dropping the presumably unimportant anisotropic stiffness term proportional to $K_A$ \cite{halperin1974analogy,de1993physics}. This transition has been extensively studied experimentally. When continuous, the exponents are either those of the $3d$ XY model, or else a slightly anisotropic version thereof \cite{de1993physics}.\footnote{Whether or not the observed anisotropy is real or simply an artefact of experimental sensitivity is a longstanding question that we will not attempt to answer, and simply refer the reader to \cite{de1993physics} for details. }
	
	
	The standard Bogoliubov treatment on the BEI side of the transition yields a massless mode that disperses quadratically at small $k$ as (setting $p=1$)
	\be \o(\bfk) = k^2\sqrt{\frac{K_D}{2\lan |\psi|^2\ran}},\ee 
	while at large $k$ the dispersion goes over to the expected $\o(\bfk) = \sqrt{\r_D} k$. Thus the hybridization with $\bfvp_D$ again ensures that the condensed phase correctly has $z=2$. 
	
	\section{Partial dipole breaking} 
	
	In this section we consider what happens when only a subset of the components of the total dipole moment are conserved. Such a scenario arises quite naturally in the context of tilted optical lattices, where partial conservation occurs if one or more principal axes of the lattice are orthogonal to the tilt direction (and is in fact the situation realized in the experiment of Ref. \cite{guardado2020subdiffusion}). The partial conservation of dipole moment allows for scenarios in which the condensed phases are insulating in some directions and superconducting in others. 
	
	\ss{Two dimensions} 
	
	In $d=2$, we consider the Hamiltonian $H = H_{hop} + H_{onsite}$, with $H_{onsite}$ the standard onsite part of the BHM (as in \eqref{ham}), and with 
	\be H_{hop} = -t_{sp} \sum_i d^x_i -t \sum_{i,a} (d^y_i)^\da d_{i+a}^y + h.c,\ee 
	which conserves only the $y$-component of the total dipole moment (in the optical lattice context, such a Hamiltonian would arise in a lattice tilted along the $y$ direction). 
	In the following we will sketch the phase diagram of this model as a function of $t/t_{sp}$, with the chemical potential and interaction strength held fixed. 
	
	Consider first the limit where the single-particle hopping vanishes, $t_{sp} = 0$. In this limit we know from previous sections what happens: at small $t$ we have a Mott insulator, at intermediate $t$ a decoupled stack of dipole condensates (with QLRO at $T=0$) stacked along the $x$ direction, and at large $t$ a decoupled stack of BEIs, with the dipole symmetry on each BEI spontaneously broken. These phases are all stable with respect to turning on a small nonzero $t_{sp}$: in the Mott insulator and dipole condensate stack the charge gap is nonzero, while in the stack of BEIs, the single-particle hopping generically acts via a perturbation to the Lagrangian of the form 
	\be \d \mcl = t_{sp} \sum_{x,m} g_m \cos(\phi_x - \phi_{x+m}),\ee 
	with the $x$ coordinate indexing the BEIs in the stack, and with $g_m$ some function decaying rapidly with $|m|$. Since $\cos(\phi_x - \phi_{x+m})$ has short-range correlations in the BEI phase, the perturbation $\d \mcl$ is irrelevant. 
	
	\begin{figure}
		\includegraphics{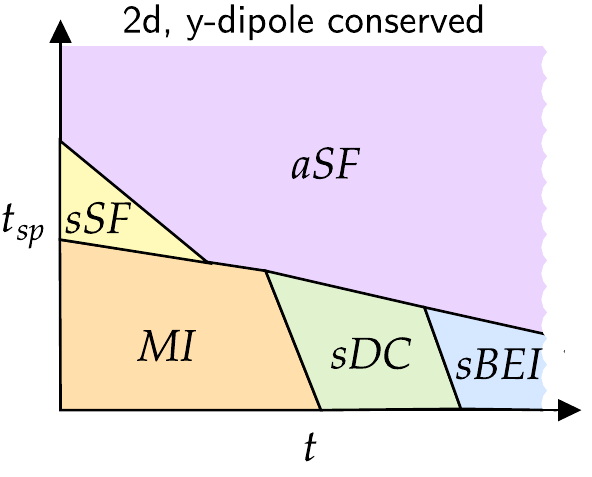}
		\caption{\label{fig:anis_2d_pd} A rough sketch of a possible phase diagram in the $t$-$t_{sp}$ plane for the 2d model that only conserves the $y$ component of the total dipole moment (with the chemical potential held fixed at some generic value). `MI' is a Mott insulator, `sDC' a stack of 1d dipole condensates, `sBEI' a stack of 1d BEIs, `sSF' a stack of 1d superfluids, and `aSF' an anisotropic superfluid phase discussed in the main text. }
	\end{figure}
	
	Now consider the limit where the dipole hopping vanishes, $t=0$. At small $t_{sp}$ we of course have a Mott insulator, while at large $t_{sp}$ we have a decoupled stack of superfluids (with QLRO at $T=0$) stacked along the $y$ direction. The Mott insulating phase is stable with respect to turning on a small nonzero $t$, while in the stack of superfluids a small $t$ acts as a perturbation like 
	\be \label{dlcosddp} \d \mcl = t\sum_{y,m,n} g_{m,n} \cos(\De^y_m \De^y_n\phi_y),\ee 
	where $\De^y_m\phi_y = \phi_{y+m}-\phi_y$ is the discrete derivative, $g_{m,n}$ decays with $|m|,|n|$, and where the subscript on $\phi$ now indexes the superfluids in the stack. Since $e^{i\phi_y}$ only has QLRO at $T=0$, the relevance of these perturbations depends on the superfluid density of the superfluids in the stack. At generic values of $\mu$ the superfluid density onsets smoothly from zero across the transition out of the Mott insulator, and there is always a regime of $t_{sp}$ for which all of the terms in $\d \mcl$ are irrelevant.\footnote{If $\mu$ is tuned so that the $t=0$ transition from the Mott insulator into the superfluid stack is of BKT type, the stiffness jumps across the transition, and the perturbations in \eqref{dlcosddp} turns out to always be relevant.} At large enough $t_{sp}$ however \eqref{dlcosddp} is relevant, so that infinitesimally small $t$ leads to an anisotropic superfluid phase described by the Lagrangian\footnote{The term $(\p_x\p_y\phi)^2$ (which is allowed by symmetry) is ignored on the grounds of it being irrelevant under a scaling for which $\tau \sim x \sim y^2$. }
	\be \label{anissf2d} \mcl = \frac\kappa2(\p_\tau\phi)^2 + \frac{K_x}2 (\p_x\phi)^2 + \frac{K_y}2(\p_y^2\phi)^2.\ee
	{It is interesting to note that the physics of this phase is quite similar to that of a quantum smectic whose layers are oriented normal to the $\uvx$ direction \cite{radzihovsky2020quantum,zhai2021fractonic}.}
	The above considerations lead to the schematic phase diagram of figure \ref{fig:anis_2d_pd}.

	By introducing a background electromagnetic field in \eqref{anissf2d}, we see that the system is superconducting along the $x$ direction and insulating along the $y$ direction. In this phase particle number conservation is spontaneously broken at $T=0$, as can be demonstrated for example by computing the correlation function 
	\bea\label{anisphicor} \lan e^{i\phi(\bfr)} e^{-i\phi(0)}\ran = \exp\( - \int \frac{d\o \, d^2k}{(\twp)^3} \frac{1-\cos(\bfk\cdot\bfr)}{\kappa\o^2 + K_x k_x^2 + K_y k_y^4}\), \eea 
	which asymptotes to a constant as $r\ra \infty$. At $T>0$ however we see that the above correlator is short-ranged, implying that particle number is not spontaneously broken. $y$-dipole conservation is spontaneously broken at $T>0$ however, as can be seen by calculating the $T>0$ correlator of $e^{i\D_y\phi}$ (done by multiplying the integrand in \eqref{anisphicor} by $(1-\cos(k_ya))^2$, with $a$ the lattice spacing).

	\ss{Three dimensions} 
	
	In $d=3$ we may consider two different scenarios. 
	
	\sss{Two components of dipole moment conserved}
	In the first scenario, two components of the dipole moment are conserved. The appropriate hopping term to study is then 
	\be H_{hop} = -t_{sp} \sum_i d^x_i - t \sum_{i,a}\( (d^y_i)^\da d^y_{i+a}+(d^z_i)^\da d^z_{i+a}\) + h.c.\ee 
	
	\begin{figure}
		\includegraphics{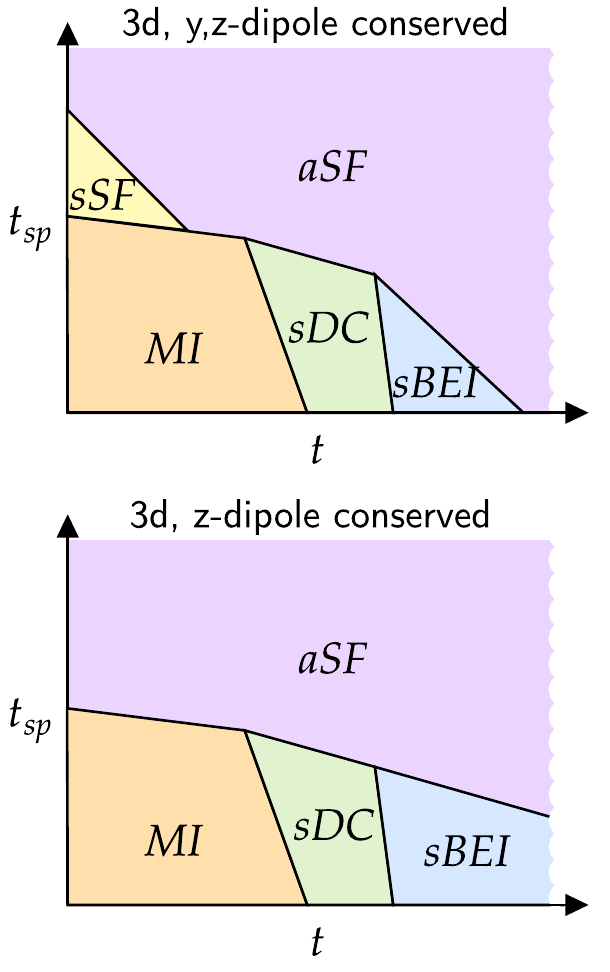}
		\caption{\label{fig:3dpd} Schematic phase diagrams for 3d models which conserve both the $y,z$ components of the total dipole moment (top) and just the $z$ component (bottom). The labeling of the various phases is as in figure \ref{fig:anis_2d_pd}.}
	\end{figure}

	The analysis of the phase diagram at small $t$ is quite similar to that of the two-dimensional model discussed in the previous section. Consider instead the limit where $t_{sp}=0$. In this limit we find the usual Mott insulator at small $t$, an array of two-dimensional dipole condensates stacked along the $x$ direction at intermediate $t$, and a stack of two-dimensional BEIs at large $t$, in which charged operators exhibit QLRO at $T=0$. The Mott insulator and dipole condensate stack are stable with respect to turning on a small $t_{sp}$, while the stability of the BEI stack depends on the relevance of the terms 
	\be\label{dl3d} \d \mcl = t_{sp} \sum_{l,m} g_m\cos(\phi_l - \phi_{l+m}),\ee 
	where $l$ indexes the different layers in the stack. The relevance of the terms in \eqref{dl3d} depends on the stiffness $K_D^{ab}$ appearing in the kinetic term $K_D^{ab}(\D_a \D_b\phi_l)^2$, with the terms in \eqref{dl3d} being irrelevant at small $K_D$ and relevant at large $K_D$. These considerations lead to the schematic phase diagram shown in the top panel of figure \ref{fig:3dpd}.
	
	The anisotropic superfluid phase where both $t,t_{sp}$ are large is described by the Lagrangian 
	\bea \label{mcl3d} \mcl & = \frac\kappa2(\p_\tau\phi)^2 + \frac{K_x}2(\p_x\phi)^2 + \frac{K_y}2(\p_y^2\phi)^2 + \frac{K_z}2(\p_z^2\phi)^2 \\ & \qq + \frac{K_{yz}}2 (\p_y\p_z\phi)^2.\eea
	In this phase the system superconducts along the $x$ direction, but is insulating along $y$ and $z$. Furthermore, by coupling \eqref{mcl3d} to a gauge field one finds that this phase exhibits {\it no} Meissner effect. This can be understood physically by recalling that the Meissner effect occurs when circulating diamagnetic supercurrents arise to cancel out an applied magnetic field. The existence of circulating supercurrents requires that the system be superconducting in more than one spatial direction, which is not the case in the present model. Finally, computations analogous to that in \eqref{anisphicor} show that total charge conservation is spontaneously broken at $T=0$ and has QLRO at $T>0$, with dipole conservation being spontaneously broken at all $T$. 
	
	\sss{One component of dipole moment conserved}
	We may also consider a scenario in which only one component of the dipole moment is conserved. In this case, we take the hopping term to be 
	\be H_{hop} = -t_{sp} \sum_{i} (d^x_i + d^y_i) - t\sum_{i,a} (d^z_i)^\da d_{i+a}^z + h.c.\ee 
	The small $t_{sp}$ portion of the phase diagram is similar to that of the $d=2$ anisotropic theory discussed previously. 
	At large $t_{sp}$, the stack of 2d superfluids that forms at $t=0$ is unstable to any finite dipole hopping at $T=0$, leading to the schematic phase diagram in the bottom panel of figure \ref{fig:3dpd}. The condensed phase at large $t,t_{sp}$ is again an anisotropic superfluid, captured by the Lagrangian 	
	\be \mcl = \frac\kappa2(\p_\tau\phi)^2 + \frac{K_x}2(\p_x\phi)^2 + \frac{K_y}2(\p_y\phi)^2 + \frac{K_z}2(\p_z^2\phi)^2.\ee 
	Thus the system superconducts along the $x$ and $y$ directions, and insulates along the $z$ direction. We also see that the system exhibits a partial Meissner effect: magnetic fields along the $z$ direction are screened, while fields along the $x$ and $y$ directions are not. As for the pattern of symmetry breaking, one finds that charge conservation is spontaneously broken for all values of $T$. 
	
	\section{Discussion}\label{sec:disc}
	
	In this paper we have explored the physics of the dipolar Bose-Hubbard model (DBHM), a simple variant of the conventional Bose-Hubbard model whose dynamics conserves both total charge and total dipole moment. The latter conservation law has a dramatic effect on the physics, significantly changing the phase diagram and producing a highly unusual phase of matter: an insulating Bose condensate, which we dub a `Bose-Einstein insulator' (BEI). We also studied models in which only some components of the total dipole moment are conserved, which were shown to produce phases that superconduct in some directions and insulate in others. 
	
	Clearly it is important to understand to what extent dipole moment conservation can be treated as a good symmetry in experimental platforms capable of simulating the DBHM. To date, the most promising experimental platforms appear to be systems of ultracold atoms prepared in tilted optical lattices \cite{khemani2020localization,guardado2020subdiffusion,scherg2021observing} (although existing experiments have only studied tilted {\it Fermi}-Hubbard models). These systems do not exactly conserve dipole moment, and any realistic microscopic Hamiltonian will possess a nonzero single-particle hopping term $-t_{sp}\sum_{i,a} b_i b^\da_{i+a} + h.c.$ However, as discussed in \cite{khemani2020localization}, in the presence of a strong tilted potential of strength $V$, these systems possess dipole-conserving dynamics over a long pre-thermal timescale $t_*$, which is exponentially large in $V/t_{sp}$, and which can even become infinite if the strength of the potential is made to scale with the system size. When $t>t_*$ the physics of the DBHM will cross over to that of a strongly tilted Bose-Hubbard model with conventional hopping terms. However, since the ground state of the dipole-conserving model has uniform density, the tilted Bose-Hubbard model describing the dynamics at $t>t_*$ will effectively be initialized at an extremely high energy state, and consequently the detailed behavior at $t>t_*$ is likely to be rather messy. Here we simply content ourselves with the fact that dipole conservation is effectively exact at large tilt potentials and in finite-sized systems, and leave a quantitative analysis of the effects of weak dipole breaking to future work. 
	
	It is also possible to consider models where the potential strength $V$ is not the largest energy scale in the problem, but is rather brought down e.g. to the level of the onsite repulsion $U$. Previous works analyzing this regime  \cite{sachdev2002mott,pielawa2011correlated} have found an interesting array of phases, with dipolar excitations remaining the star of the show. It would be interesting to understand how this physics connects to that of the $V\gg U$ regime considered in this paper.

	One of the most interesting aspects of systems whose dynamics conserves dipole moment is the strong sensitivity of the system's dynamics to the choice of initial state, with it often being the case that a large fraction of the Hilbert space is completely inert under time evolution \cite{khemani2020localization}.
	This however is not likely to be an issue for performing an experimental study of the universal aspects of the equilibrium DBHM phase diagram.
	This is so because there is always a canonical choice for the initial state prepared in experiments, which belongs to an exponentially large sector of Hilbert space in which the dipolar dynamics act ergodically, in contrast to an inert ``shattered'' state. For example, in optical lattice realizations, we can imagine first preparing the system in an un-tilted lattice with weak intersite tunneling, placing the system deep in a Mott insulating phase. We can then turn on the tilt potential while remaining in the Mott insulator, and from here one can subsequently increase the tunneling and study the phases that occur at larger hopping strengths. This procedure sidesteps issues of non-standard thermalization due to Hilbert space shattering.
	
	\section*{Acknowledgments} 
	
	We thank David Huse, Kristan Jensen, Ho Tat Lam, and especially Leo Radzihovsky for discussions. EL is supported by the Hertz Fellowship. The research of MH is supported by the U.S. Department of Energy, Office of Science, Basic Energy Sciences (BES) under Award number DE-SC0014415. TS was supported by US Department of Energy grant DE-SC0008739, and partially through a Simons Investigator Award from the Simons Foundation. This work was also partly supported by the Simons Collaboration on Ultra-Quantum Matter, which is a grant from the Simons Foundation (651440, MH; 651446, TS).

	\bibliography{dbhm_bib}

\begin{thebibliography}{52}
\expandafter\ifx\csname natexlab\endcsname\relax\def\natexlab#1{#1}\fi
\expandafter\ifx\csname bibnamefont\endcsname\relax
  \def\bibnamefont#1{#1}\fi
\expandafter\ifx\csname bibfnamefont\endcsname\relax
  \def\bibfnamefont#1{#1}\fi
\expandafter\ifx\csname citenamefont\endcsname\relax
  \def\citenamefont#1{#1}\fi
\expandafter\ifx\csname url\endcsname\relax
  \def\url#1{\texttt{#1}}\fi
\expandafter\ifx\csname urlprefix\endcsname\relax\def\urlprefix{URL }\fi
\providecommand{\bibinfo}[2]{#2}
\providecommand{\eprint}[2][]{\url{#2}}

\bibitem[{\citenamefont{Pretko}(2017)}]{pretko2017subdimensional}
\bibinfo{author}{\bibfnamefont{M.}~\bibnamefont{Pretko}},
  \bibinfo{journal}{Physical Review B} \textbf{\bibinfo{volume}{95}},
  \bibinfo{pages}{115139} (\bibinfo{year}{2017}).

\bibitem[{\citenamefont{Pretko}(2018)}]{pretko2018fracton}
\bibinfo{author}{\bibfnamefont{M.}~\bibnamefont{Pretko}},
  \bibinfo{journal}{Physical Review B} \textbf{\bibinfo{volume}{98}},
  \bibinfo{pages}{115134} (\bibinfo{year}{2018}).

\bibitem[{\citenamefont{Nandkishore and
  Hermele}(2019)}]{nandkishore2019fractons}
\bibinfo{author}{\bibfnamefont{R.~M.} \bibnamefont{Nandkishore}}
  \bibnamefont{and} \bibinfo{author}{\bibfnamefont{M.}~\bibnamefont{Hermele}},
  \bibinfo{journal}{Annual Review of Condensed Matter Physics}
  \textbf{\bibinfo{volume}{10}}, \bibinfo{pages}{295} (\bibinfo{year}{2019}).

\bibitem[{\citenamefont{Seiberg}(2020)}]{seiberg2020field}
\bibinfo{author}{\bibfnamefont{N.}~\bibnamefont{Seiberg}},
  \bibinfo{journal}{arXiv preprint arXiv:1909.10544}  (\bibinfo{year}{2020}).

\bibitem[{\citenamefont{Pretko et~al.}(2020)\citenamefont{Pretko, Chen, and
  You}}]{pretko2020fracton}
\bibinfo{author}{\bibfnamefont{M.}~\bibnamefont{Pretko}},
  \bibinfo{author}{\bibfnamefont{X.}~\bibnamefont{Chen}}, \bibnamefont{and}
  \bibinfo{author}{\bibfnamefont{Y.}~\bibnamefont{You}},
  \bibinfo{journal}{International Journal of Modern Physics A}
  \textbf{\bibinfo{volume}{35}}, \bibinfo{pages}{2030003}
  (\bibinfo{year}{2020}).

\bibitem[{\citenamefont{Radzihovsky and
  Hermele}(2020)}]{radzihovsky2020fractons}
\bibinfo{author}{\bibfnamefont{L.}~\bibnamefont{Radzihovsky}} \bibnamefont{and}
  \bibinfo{author}{\bibfnamefont{M.}~\bibnamefont{Hermele}},
  \bibinfo{journal}{Physical review letters} \textbf{\bibinfo{volume}{124}},
  \bibinfo{pages}{050402} (\bibinfo{year}{2020}).

\bibitem[{\citenamefont{Griffin et~al.}(2015)\citenamefont{Griffin, Grosvenor,
  Ho{\v{r}}ava, and Yan}}]{griffin2015scalar}
\bibinfo{author}{\bibfnamefont{T.}~\bibnamefont{Griffin}},
  \bibinfo{author}{\bibfnamefont{K.~T.} \bibnamefont{Grosvenor}},
  \bibinfo{author}{\bibfnamefont{P.}~\bibnamefont{Ho{\v{r}}ava}},
  \bibnamefont{and} \bibinfo{author}{\bibfnamefont{Z.}~\bibnamefont{Yan}},
  \bibinfo{journal}{Communications in Mathematical Physics}
  \textbf{\bibinfo{volume}{340}}, \bibinfo{pages}{985} (\bibinfo{year}{2015}).

\bibitem[{\citenamefont{Bidussi et~al.}(2021)\citenamefont{Bidussi, Hartong,
  Have, Musaeus, and Prohazka}}]{bidussi2021fractons}
\bibinfo{author}{\bibfnamefont{L.}~\bibnamefont{Bidussi}},
  \bibinfo{author}{\bibfnamefont{J.}~\bibnamefont{Hartong}},
  \bibinfo{author}{\bibfnamefont{E.}~\bibnamefont{Have}},
  \bibinfo{author}{\bibfnamefont{J.}~\bibnamefont{Musaeus}}, \bibnamefont{and}
  \bibinfo{author}{\bibfnamefont{S.}~\bibnamefont{Prohazka}},
  \bibinfo{journal}{arXiv preprint arXiv:2111.03668}  (\bibinfo{year}{2021}).

\bibitem[{\citenamefont{Jain and Jensen}(2021)}]{jain2021fractons}
\bibinfo{author}{\bibfnamefont{A.}~\bibnamefont{Jain}} \bibnamefont{and}
  \bibinfo{author}{\bibfnamefont{K.}~\bibnamefont{Jensen}},
  \bibinfo{journal}{arXiv preprint arXiv:2111.03973}  (\bibinfo{year}{2021}).

\bibitem[{\citenamefont{Grosvenor
  et~al.}(2021{\natexlab{a}})\citenamefont{Grosvenor, Hoyos, Pe{\~n}a-Benitez,
  and Sur{\'o}wka}}]{grosvenor2021space}
\bibinfo{author}{\bibfnamefont{K.~T.} \bibnamefont{Grosvenor}},
  \bibinfo{author}{\bibfnamefont{C.}~\bibnamefont{Hoyos}},
  \bibinfo{author}{\bibfnamefont{F.}~\bibnamefont{Pe{\~n}a-Benitez}},
  \bibnamefont{and}
  \bibinfo{author}{\bibfnamefont{P.}~\bibnamefont{Sur{\'o}wka}},
  \bibinfo{journal}{arXiv preprint arXiv:2112.00531}
  (\bibinfo{year}{2021}{\natexlab{a}}).

\bibitem[{\citenamefont{Stahl et~al.}(2021)\citenamefont{Stahl, Lake, and
  Nandkishore}}]{stahl2021spontaneous}
\bibinfo{author}{\bibfnamefont{C.}~\bibnamefont{Stahl}},
  \bibinfo{author}{\bibfnamefont{E.}~\bibnamefont{Lake}}, \bibnamefont{and}
  \bibinfo{author}{\bibfnamefont{R.}~\bibnamefont{Nandkishore}},
  \bibinfo{journal}{arXiv preprint arXiv:2111.08041}  (\bibinfo{year}{2021}).

\bibitem[{\citenamefont{Pai et~al.}(2019)\citenamefont{Pai, Pretko, and
  Nandkishore}}]{pai2019localization}
\bibinfo{author}{\bibfnamefont{S.}~\bibnamefont{Pai}},
  \bibinfo{author}{\bibfnamefont{M.}~\bibnamefont{Pretko}}, \bibnamefont{and}
  \bibinfo{author}{\bibfnamefont{R.~M.} \bibnamefont{Nandkishore}},
  \bibinfo{journal}{Physical Review X} \textbf{\bibinfo{volume}{9}},
  \bibinfo{pages}{021003} (\bibinfo{year}{2019}).

\bibitem[{\citenamefont{Khemani et~al.}(2020)\citenamefont{Khemani, Hermele,
  and Nandkishore}}]{khemani2020localization}
\bibinfo{author}{\bibfnamefont{V.}~\bibnamefont{Khemani}},
  \bibinfo{author}{\bibfnamefont{M.}~\bibnamefont{Hermele}}, \bibnamefont{and}
  \bibinfo{author}{\bibfnamefont{R.}~\bibnamefont{Nandkishore}},
  \bibinfo{journal}{Physical Review B} \textbf{\bibinfo{volume}{101}},
  \bibinfo{pages}{174204} (\bibinfo{year}{2020}).

\bibitem[{\citenamefont{Sala et~al.}(2020)\citenamefont{Sala, Rakovszky,
  Verresen, Knap, and Pollmann}}]{sala2020ergodicity}
\bibinfo{author}{\bibfnamefont{P.}~\bibnamefont{Sala}},
  \bibinfo{author}{\bibfnamefont{T.}~\bibnamefont{Rakovszky}},
  \bibinfo{author}{\bibfnamefont{R.}~\bibnamefont{Verresen}},
  \bibinfo{author}{\bibfnamefont{M.}~\bibnamefont{Knap}}, \bibnamefont{and}
  \bibinfo{author}{\bibfnamefont{F.}~\bibnamefont{Pollmann}},
  \bibinfo{journal}{Physical Review X} \textbf{\bibinfo{volume}{10}},
  \bibinfo{pages}{011047} (\bibinfo{year}{2020}).

\bibitem[{\citenamefont{Rakovszky et~al.}(2020)\citenamefont{Rakovszky, Sala,
  Verresen, Knap, and Pollmann}}]{rakovszky2020statistical}
\bibinfo{author}{\bibfnamefont{T.}~\bibnamefont{Rakovszky}},
  \bibinfo{author}{\bibfnamefont{P.}~\bibnamefont{Sala}},
  \bibinfo{author}{\bibfnamefont{R.}~\bibnamefont{Verresen}},
  \bibinfo{author}{\bibfnamefont{M.}~\bibnamefont{Knap}}, \bibnamefont{and}
  \bibinfo{author}{\bibfnamefont{F.}~\bibnamefont{Pollmann}},
  \bibinfo{journal}{Physical Review B} \textbf{\bibinfo{volume}{101}},
  \bibinfo{pages}{125126} (\bibinfo{year}{2020}).

\bibitem[{\citenamefont{Moudgalya et~al.}(2019)\citenamefont{Moudgalya, Prem,
  Nandkishore, Regnault, and Bernevig}}]{moudgalya2019thermalization}
\bibinfo{author}{\bibfnamefont{S.}~\bibnamefont{Moudgalya}},
  \bibinfo{author}{\bibfnamefont{A.}~\bibnamefont{Prem}},
  \bibinfo{author}{\bibfnamefont{R.}~\bibnamefont{Nandkishore}},
  \bibinfo{author}{\bibfnamefont{N.}~\bibnamefont{Regnault}}, \bibnamefont{and}
  \bibinfo{author}{\bibfnamefont{B.~A.} \bibnamefont{Bernevig}},
  \bibinfo{journal}{arXiv preprint arXiv:1910.14048}  (\bibinfo{year}{2019}).

\bibitem[{\citenamefont{Gromov et~al.}(2020)\citenamefont{Gromov, Lucas, and
  Nandkishore}}]{gromov2020fracton}
\bibinfo{author}{\bibfnamefont{A.}~\bibnamefont{Gromov}},
  \bibinfo{author}{\bibfnamefont{A.}~\bibnamefont{Lucas}}, \bibnamefont{and}
  \bibinfo{author}{\bibfnamefont{R.~M.} \bibnamefont{Nandkishore}},
  \bibinfo{journal}{Physical Review Research} \textbf{\bibinfo{volume}{2}},
  \bibinfo{pages}{033124} (\bibinfo{year}{2020}).

\bibitem[{\citenamefont{Feldmeier et~al.}(2020)\citenamefont{Feldmeier, Sala,
  De~Tomasi, Pollmann, and Knap}}]{feldmeier2020anomalous}
\bibinfo{author}{\bibfnamefont{J.}~\bibnamefont{Feldmeier}},
  \bibinfo{author}{\bibfnamefont{P.}~\bibnamefont{Sala}},
  \bibinfo{author}{\bibfnamefont{G.}~\bibnamefont{De~Tomasi}},
  \bibinfo{author}{\bibfnamefont{F.}~\bibnamefont{Pollmann}}, \bibnamefont{and}
  \bibinfo{author}{\bibfnamefont{M.}~\bibnamefont{Knap}},
  \bibinfo{journal}{Physical Review Letters} \textbf{\bibinfo{volume}{125}},
  \bibinfo{pages}{245303} (\bibinfo{year}{2020}).

\bibitem[{\citenamefont{Iaconis et~al.}(2021)\citenamefont{Iaconis, Lucas, and
  Nandkishore}}]{iaconis2021multipole}
\bibinfo{author}{\bibfnamefont{J.}~\bibnamefont{Iaconis}},
  \bibinfo{author}{\bibfnamefont{A.}~\bibnamefont{Lucas}}, \bibnamefont{and}
  \bibinfo{author}{\bibfnamefont{R.}~\bibnamefont{Nandkishore}},
  \bibinfo{journal}{Physical Review E} \textbf{\bibinfo{volume}{103}},
  \bibinfo{pages}{022142} (\bibinfo{year}{2021}).

\bibitem[{\citenamefont{Glorioso et~al.}(2021)\citenamefont{Glorioso, Guo,
  Rodriguez-Nieva, and Lucas}}]{glorioso2021breakdown}
\bibinfo{author}{\bibfnamefont{P.}~\bibnamefont{Glorioso}},
  \bibinfo{author}{\bibfnamefont{J.}~\bibnamefont{Guo}},
  \bibinfo{author}{\bibfnamefont{J.~F.} \bibnamefont{Rodriguez-Nieva}},
  \bibnamefont{and} \bibinfo{author}{\bibfnamefont{A.}~\bibnamefont{Lucas}},
  \bibinfo{journal}{arXiv preprint arXiv:2105.13365}  (\bibinfo{year}{2021}).

\bibitem[{\citenamefont{Grosvenor
  et~al.}(2021{\natexlab{b}})\citenamefont{Grosvenor, Hoyos, Pe{\~n}a-Benitez,
  and Sur{\'o}wka}}]{grosvenor2021hydrodynamics}
\bibinfo{author}{\bibfnamefont{K.~T.} \bibnamefont{Grosvenor}},
  \bibinfo{author}{\bibfnamefont{C.}~\bibnamefont{Hoyos}},
  \bibinfo{author}{\bibfnamefont{F.}~\bibnamefont{Pe{\~n}a-Benitez}},
  \bibnamefont{and}
  \bibinfo{author}{\bibfnamefont{P.}~\bibnamefont{Sur{\'o}wka}},
  \bibinfo{journal}{arXiv preprint arXiv:2105.01084}
  (\bibinfo{year}{2021}{\natexlab{b}}).

\bibitem[{\citenamefont{Radzihovsky}(2020)}]{radzihovsky2020quantum}
\bibinfo{author}{\bibfnamefont{L.}~\bibnamefont{Radzihovsky}},
  \bibinfo{journal}{Physical Review Letters} \textbf{\bibinfo{volume}{125}},
  \bibinfo{pages}{267601} (\bibinfo{year}{2020}).

\bibitem[{\citenamefont{Guardado-Sanchez
  et~al.}(2020)\citenamefont{Guardado-Sanchez, Morningstar, Spar, Brown, Huse,
  and Bakr}}]{guardado2020subdiffusion}
\bibinfo{author}{\bibfnamefont{E.}~\bibnamefont{Guardado-Sanchez}},
  \bibinfo{author}{\bibfnamefont{A.}~\bibnamefont{Morningstar}},
  \bibinfo{author}{\bibfnamefont{B.~M.} \bibnamefont{Spar}},
  \bibinfo{author}{\bibfnamefont{P.~T.} \bibnamefont{Brown}},
  \bibinfo{author}{\bibfnamefont{D.~A.} \bibnamefont{Huse}}, \bibnamefont{and}
  \bibinfo{author}{\bibfnamefont{W.~S.} \bibnamefont{Bakr}},
  \bibinfo{journal}{Physical Review X} \textbf{\bibinfo{volume}{10}},
  \bibinfo{pages}{011042} (\bibinfo{year}{2020}).

\bibitem[{\citenamefont{Scherg et~al.}(2021)\citenamefont{Scherg, Kohlert,
  Sala, Pollmann, Madhusudhana, Bloch, and Aidelsburger}}]{scherg2021observing}
\bibinfo{author}{\bibfnamefont{S.}~\bibnamefont{Scherg}},
  \bibinfo{author}{\bibfnamefont{T.}~\bibnamefont{Kohlert}},
  \bibinfo{author}{\bibfnamefont{P.}~\bibnamefont{Sala}},
  \bibinfo{author}{\bibfnamefont{F.}~\bibnamefont{Pollmann}},
  \bibinfo{author}{\bibfnamefont{B.~H.} \bibnamefont{Madhusudhana}},
  \bibinfo{author}{\bibfnamefont{I.}~\bibnamefont{Bloch}}, \bibnamefont{and}
  \bibinfo{author}{\bibfnamefont{M.}~\bibnamefont{Aidelsburger}},
  \bibinfo{journal}{Nature Communications} \textbf{\bibinfo{volume}{12}},
  \bibinfo{pages}{1} (\bibinfo{year}{2021}).

\bibitem[{\citenamefont{Kohlert et~al.}(2021)\citenamefont{Kohlert, Scherg,
  Sala, Pollmann, Madhusudhana, Bloch, and
  Aidelsburger}}]{kohlert2021experimental}
\bibinfo{author}{\bibfnamefont{T.}~\bibnamefont{Kohlert}},
  \bibinfo{author}{\bibfnamefont{S.}~\bibnamefont{Scherg}},
  \bibinfo{author}{\bibfnamefont{P.}~\bibnamefont{Sala}},
  \bibinfo{author}{\bibfnamefont{F.}~\bibnamefont{Pollmann}},
  \bibinfo{author}{\bibfnamefont{B.~H.} \bibnamefont{Madhusudhana}},
  \bibinfo{author}{\bibfnamefont{I.}~\bibnamefont{Bloch}}, \bibnamefont{and}
  \bibinfo{author}{\bibfnamefont{M.}~\bibnamefont{Aidelsburger}},
  \bibinfo{journal}{arXiv preprint arXiv:2106.15586}  (\bibinfo{year}{2021}).

\bibitem[{\citenamefont{Fisher et~al.}(1989)\citenamefont{Fisher, Weichman,
  Grinstein, and Fisher}}]{fisher1989boson}
\bibinfo{author}{\bibfnamefont{M.~P.} \bibnamefont{Fisher}},
  \bibinfo{author}{\bibfnamefont{P.~B.} \bibnamefont{Weichman}},
  \bibinfo{author}{\bibfnamefont{G.}~\bibnamefont{Grinstein}},
  \bibnamefont{and} \bibinfo{author}{\bibfnamefont{D.~S.}
  \bibnamefont{Fisher}}, \bibinfo{journal}{Physical Review B}
  \textbf{\bibinfo{volume}{40}}, \bibinfo{pages}{546} (\bibinfo{year}{1989}).

\bibitem[{\citenamefont{Sachdev}(2011)}]{sachdev2011quantum}
\bibinfo{author}{\bibfnamefont{S.}~\bibnamefont{Sachdev}},
  \emph{\bibinfo{title}{Quantum phase transitions}}
  (\bibinfo{publisher}{Cambridge university press}, \bibinfo{year}{2011}).

\bibitem[{\citenamefont{Prem et~al.}(2018)\citenamefont{Prem, Pretko, and
  Nandkishore}}]{prem2018emergent}
\bibinfo{author}{\bibfnamefont{A.}~\bibnamefont{Prem}},
  \bibinfo{author}{\bibfnamefont{M.}~\bibnamefont{Pretko}}, \bibnamefont{and}
  \bibinfo{author}{\bibfnamefont{R.~M.} \bibnamefont{Nandkishore}},
  \bibinfo{journal}{Physical Review B} \textbf{\bibinfo{volume}{97}},
  \bibinfo{pages}{085116} (\bibinfo{year}{2018}).

\bibitem[{\citenamefont{et~al.}(2022)}]{lake20221dmodel}
\bibinfo{author}{\bibfnamefont{L.}~\bibnamefont{et~al.}}, \bibinfo{journal}{To
  appear}  (\bibinfo{year}{2022}).

\bibitem[{\citenamefont{Kumar and Potter}(2019)}]{kumar2019symmetry}
\bibinfo{author}{\bibfnamefont{A.}~\bibnamefont{Kumar}} \bibnamefont{and}
  \bibinfo{author}{\bibfnamefont{A.~C.} \bibnamefont{Potter}},
  \bibinfo{journal}{Physical Review B} \textbf{\bibinfo{volume}{100}},
  \bibinfo{pages}{045119} (\bibinfo{year}{2019}).

\bibitem[{\citenamefont{Pretko and Radzihovsky}(2018)}]{pretko2018symmetry}
\bibinfo{author}{\bibfnamefont{M.}~\bibnamefont{Pretko}} \bibnamefont{and}
  \bibinfo{author}{\bibfnamefont{L.}~\bibnamefont{Radzihovsky}},
  \bibinfo{journal}{Physical review letters} \textbf{\bibinfo{volume}{121}},
  \bibinfo{pages}{235301} (\bibinfo{year}{2018}).

\bibitem[{\citenamefont{Zhai and Radzihovsky}(2021)}]{zhai2021fractonic}
\bibinfo{author}{\bibfnamefont{Z.}~\bibnamefont{Zhai}} \bibnamefont{and}
  \bibinfo{author}{\bibfnamefont{L.}~\bibnamefont{Radzihovsky}},
  \bibinfo{journal}{Annals of Physics} p. \bibinfo{pages}{168509}
  (\bibinfo{year}{2021}).

\bibitem[{\citenamefont{Yuan et~al.}(2020)\citenamefont{Yuan, Chen, and
  Ye}}]{yuan2020fractonic}
\bibinfo{author}{\bibfnamefont{J.-K.} \bibnamefont{Yuan}},
  \bibinfo{author}{\bibfnamefont{S.~A.} \bibnamefont{Chen}}, \bibnamefont{and}
  \bibinfo{author}{\bibfnamefont{P.}~\bibnamefont{Ye}},
  \bibinfo{journal}{Physical Review Research} \textbf{\bibinfo{volume}{2}},
  \bibinfo{pages}{023267} (\bibinfo{year}{2020}).

\bibitem[{\citenamefont{Chen et~al.}(2021)\citenamefont{Chen, Yuan, and
  Ye}}]{chen2021fractonic}
\bibinfo{author}{\bibfnamefont{S.~A.} \bibnamefont{Chen}},
  \bibinfo{author}{\bibfnamefont{J.-K.} \bibnamefont{Yuan}}, \bibnamefont{and}
  \bibinfo{author}{\bibfnamefont{P.}~\bibnamefont{Ye}},
  \bibinfo{journal}{Physical Review Research} \textbf{\bibinfo{volume}{3}},
  \bibinfo{pages}{013226} (\bibinfo{year}{2021}).

\bibitem[{\citenamefont{Gorantla et~al.}(2022)\citenamefont{Gorantla, Lam,
  Seiberg, and Shao}}]{gorantla2022global}
\bibinfo{author}{\bibfnamefont{P.}~\bibnamefont{Gorantla}},
  \bibinfo{author}{\bibfnamefont{H.~T.} \bibnamefont{Lam}},
  \bibinfo{author}{\bibfnamefont{N.}~\bibnamefont{Seiberg}}, \bibnamefont{and}
  \bibinfo{author}{\bibfnamefont{S.-H.} \bibnamefont{Shao}},
  \bibinfo{journal}{arXiv preprint arXiv:2201.10589}  (\bibinfo{year}{2022}).

\bibitem[{\citenamefont{Fradkin et~al.}(2004)\citenamefont{Fradkin, Huse,
  Moessner, Oganesyan, and Sondhi}}]{fradkin2004bipartite}
\bibinfo{author}{\bibfnamefont{E.}~\bibnamefont{Fradkin}},
  \bibinfo{author}{\bibfnamefont{D.~A.} \bibnamefont{Huse}},
  \bibinfo{author}{\bibfnamefont{R.}~\bibnamefont{Moessner}},
  \bibinfo{author}{\bibfnamefont{V.}~\bibnamefont{Oganesyan}},
  \bibnamefont{and} \bibinfo{author}{\bibfnamefont{S.~L.}
  \bibnamefont{Sondhi}}, \bibinfo{journal}{Physical Review B}
  \textbf{\bibinfo{volume}{69}}, \bibinfo{pages}{224415}
  (\bibinfo{year}{2004}).

\bibitem[{\citenamefont{Zhang et~al.}(2016)\citenamefont{Zhang, Yu, Ng, Zhang,
  Pitaevskii, and Stringari}}]{zhang2016superfluid}
\bibinfo{author}{\bibfnamefont{Y.-C.} \bibnamefont{Zhang}},
  \bibinfo{author}{\bibfnamefont{Z.-Q.} \bibnamefont{Yu}},
  \bibinfo{author}{\bibfnamefont{T.~K.} \bibnamefont{Ng}},
  \bibinfo{author}{\bibfnamefont{S.}~\bibnamefont{Zhang}},
  \bibinfo{author}{\bibfnamefont{L.}~\bibnamefont{Pitaevskii}},
  \bibnamefont{and}
  \bibinfo{author}{\bibfnamefont{S.}~\bibnamefont{Stringari}},
  \bibinfo{journal}{Physical Review A} \textbf{\bibinfo{volume}{94}},
  \bibinfo{pages}{033635} (\bibinfo{year}{2016}).

\bibitem[{\citenamefont{Lake and Senthil}(2021)}]{lake2021re}
\bibinfo{author}{\bibfnamefont{E.}~\bibnamefont{Lake}} \bibnamefont{and}
  \bibinfo{author}{\bibfnamefont{T.}~\bibnamefont{Senthil}},
  \bibinfo{journal}{Phys. Rev. B} \textbf{\bibinfo{volume}{104}},
  \bibinfo{pages}{174505} (\bibinfo{year}{2021}).

\bibitem[{\citenamefont{Ma and Pretko}(2018)}]{ma2018higher}
\bibinfo{author}{\bibfnamefont{H.}~\bibnamefont{Ma}} \bibnamefont{and}
  \bibinfo{author}{\bibfnamefont{M.}~\bibnamefont{Pretko}},
  \bibinfo{journal}{Physical Review B} \textbf{\bibinfo{volume}{98}},
  \bibinfo{pages}{125105} (\bibinfo{year}{2018}).

\bibitem[{\citenamefont{Vishwanath et~al.}(2004)\citenamefont{Vishwanath,
  Balents, and Senthil}}]{vishwanath2004quantum}
\bibinfo{author}{\bibfnamefont{A.}~\bibnamefont{Vishwanath}},
  \bibinfo{author}{\bibfnamefont{L.}~\bibnamefont{Balents}}, \bibnamefont{and}
  \bibinfo{author}{\bibfnamefont{T.}~\bibnamefont{Senthil}},
  \bibinfo{journal}{Physical Review B} \textbf{\bibinfo{volume}{69}},
  \bibinfo{pages}{224416} (\bibinfo{year}{2004}).

\bibitem[{\citenamefont{Ghaemi et~al.}(2005)\citenamefont{Ghaemi, Vishwanath,
  and Senthil}}]{ghaemi2005finite}
\bibinfo{author}{\bibfnamefont{P.}~\bibnamefont{Ghaemi}},
  \bibinfo{author}{\bibfnamefont{A.}~\bibnamefont{Vishwanath}},
  \bibnamefont{and} \bibinfo{author}{\bibfnamefont{T.}~\bibnamefont{Senthil}},
  \bibinfo{journal}{Physical Review B} \textbf{\bibinfo{volume}{72}},
  \bibinfo{pages}{024420} (\bibinfo{year}{2005}).

\bibitem[{\citenamefont{Else and Senthil}(2021)}]{else2021critical}
\bibinfo{author}{\bibfnamefont{D.}~\bibnamefont{Else}} \bibnamefont{and}
  \bibinfo{author}{\bibfnamefont{T.}~\bibnamefont{Senthil}},
  \bibinfo{journal}{arXiv preprint arXiv:2106.15623}  (\bibinfo{year}{2021}).

\bibitem[{\citenamefont{Calabrese et~al.}(2004)\citenamefont{Calabrese,
  Parruccini, Pelissetto, and Vicari}}]{calabrese2004critical}
\bibinfo{author}{\bibfnamefont{P.}~\bibnamefont{Calabrese}},
  \bibinfo{author}{\bibfnamefont{P.}~\bibnamefont{Parruccini}},
  \bibinfo{author}{\bibfnamefont{A.}~\bibnamefont{Pelissetto}},
  \bibnamefont{and} \bibinfo{author}{\bibfnamefont{E.}~\bibnamefont{Vicari}},
  \bibinfo{journal}{Physical Review B} \textbf{\bibinfo{volume}{70}},
  \bibinfo{pages}{174439} (\bibinfo{year}{2004}).

\bibitem[{\citenamefont{Poland et~al.}(2019)\citenamefont{Poland, Rychkov, and
  Vichi}}]{poland2019conformal}
\bibinfo{author}{\bibfnamefont{D.}~\bibnamefont{Poland}},
  \bibinfo{author}{\bibfnamefont{S.}~\bibnamefont{Rychkov}}, \bibnamefont{and}
  \bibinfo{author}{\bibfnamefont{A.}~\bibnamefont{Vichi}},
  \bibinfo{journal}{Reviews of Modern Physics} \textbf{\bibinfo{volume}{91}},
  \bibinfo{pages}{015002} (\bibinfo{year}{2019}).

\bibitem[{\citenamefont{Coleman and Weinberg}(1973)}]{coleman1973radiative}
\bibinfo{author}{\bibfnamefont{S.}~\bibnamefont{Coleman}} \bibnamefont{and}
  \bibinfo{author}{\bibfnamefont{E.}~\bibnamefont{Weinberg}},
  \bibinfo{journal}{Physical Review D} \textbf{\bibinfo{volume}{7}},
  \bibinfo{pages}{1888} (\bibinfo{year}{1973}).

\bibitem[{\citenamefont{Halperin et~al.}(1974)\citenamefont{Halperin, Lubensky,
  and Ma}}]{halperin1974first}
\bibinfo{author}{\bibfnamefont{B.}~\bibnamefont{Halperin}},
  \bibinfo{author}{\bibfnamefont{T.}~\bibnamefont{Lubensky}}, \bibnamefont{and}
  \bibinfo{author}{\bibfnamefont{S.-k.} \bibnamefont{Ma}},
  \bibinfo{journal}{Physical Review Letters} \textbf{\bibinfo{volume}{32}},
  \bibinfo{pages}{292} (\bibinfo{year}{1974}).

\bibitem[{\citenamefont{Halperin and Lubensky}(1974)}]{halperin1974analogy}
\bibinfo{author}{\bibfnamefont{B.}~\bibnamefont{Halperin}} \bibnamefont{and}
  \bibinfo{author}{\bibfnamefont{T.}~\bibnamefont{Lubensky}},
  \bibinfo{journal}{Solid State Communications} \textbf{\bibinfo{volume}{14}},
  \bibinfo{pages}{997} (\bibinfo{year}{1974}).

\bibitem[{\citenamefont{De~Gennes and Prost}(1993)}]{de1993physics}
\bibinfo{author}{\bibfnamefont{P.-G.} \bibnamefont{De~Gennes}}
  \bibnamefont{and} \bibinfo{author}{\bibfnamefont{J.}~\bibnamefont{Prost}},
  \emph{\bibinfo{title}{The physics of liquid crystals}}, \bibinfo{number}{83}
  (\bibinfo{publisher}{Oxford university press}, \bibinfo{year}{1993}).

\bibitem[{\citenamefont{Sachdev et~al.}(2002)\citenamefont{Sachdev, Sengupta,
  and Girvin}}]{sachdev2002mott}
\bibinfo{author}{\bibfnamefont{S.}~\bibnamefont{Sachdev}},
  \bibinfo{author}{\bibfnamefont{K.}~\bibnamefont{Sengupta}}, \bibnamefont{and}
  \bibinfo{author}{\bibfnamefont{S.}~\bibnamefont{Girvin}},
  \bibinfo{journal}{Physical Review B} \textbf{\bibinfo{volume}{66}},
  \bibinfo{pages}{075128} (\bibinfo{year}{2002}).

\bibitem[{\citenamefont{Pielawa et~al.}(2011)\citenamefont{Pielawa, Kitagawa,
  Berg, and Sachdev}}]{pielawa2011correlated}
\bibinfo{author}{\bibfnamefont{S.}~\bibnamefont{Pielawa}},
  \bibinfo{author}{\bibfnamefont{T.}~\bibnamefont{Kitagawa}},
  \bibinfo{author}{\bibfnamefont{E.}~\bibnamefont{Berg}}, \bibnamefont{and}
  \bibinfo{author}{\bibfnamefont{S.}~\bibnamefont{Sachdev}},
  \bibinfo{journal}{Physical Review B} \textbf{\bibinfo{volume}{83}},
  \bibinfo{pages}{205135} (\bibinfo{year}{2011}).

\bibitem[{\citenamefont{Lake and Hermele}(2021)}]{lake2021subdimensional}
\bibinfo{author}{\bibfnamefont{E.}~\bibnamefont{Lake}} \bibnamefont{and}
  \bibinfo{author}{\bibfnamefont{M.}~\bibnamefont{Hermele}},
  \bibinfo{journal}{Physical Review B} \textbf{\bibinfo{volume}{104}},
  \bibinfo{pages}{165121} (\bibinfo{year}{2021}).

\bibitem[{\citenamefont{Lake}(2021)}]{lake2021rg}
\bibinfo{author}{\bibfnamefont{E.}~\bibnamefont{Lake}}, \bibinfo{journal}{arXiv
  preprint arXiv:2110.02986}  (\bibinfo{year}{2021}).

\end{thebibliography}
	
	\appendix 
	
	\begin{widetext} 
		
		\section{Single dipole condensate in $d>1$}\label{app:single_dipole_cond}
		
		In this appendix we consider what happens in $d>1$ dimensions when interactions favor a scenario in which only a single species of dipole moment condenses (which we take without loss of generality to be $d^x$). 
		
		The phase where $d^x$ dipoles have condensed but individual bosons are gapped is described simply by a single compact scalar $\vp$, the phase mode of the dipole condensate. More interesting is the regime in which individual bosons are condensed.
		Because individual bosons in the dipole condensed phase only possess an effective single-particle hopping along the $\uvx$ direction, we may analyze this regime by way of a quasi-one-dimensional description in terms of a coupled array of Luttinger liquids, in a manner quite similar to the subdimensional critical points considered in \cite{lake2021subdimensional}. 
		
		We will index the Luttinger liquids by a $(d-1)$-dimensional vector $\bfl$, which runs over the sites of a $(d-1)$-dimensional square lattice. Writing the phase field for the Luttinger liquid at site $\bfl$ as $\phi_\bfl$, we then define fields Fourier transformed in the $d-1$ directions normal to $\uvx$ as $\phi_\bfp(x,\tau) = \sum_\bfl e^{-i\bfl\cdot\bfp} \phi_\bfl(x,\tau)$. The most general action we may write down for the $\phi_\bfp,\vp_\bfp$ fields is 
		\bea S =  \int d\tau\, dx\, \Bigg(\frac12 \int\frac{d^{d-1}p}{(\twp)^{d-1}} \Big( \mcj_\bfp|\p_\tau\phi_\bfp|^2 + \mck_\bfp|\p_x\phi_\bfp-\vp_\bfp|^2  + \mcm_\bfp|\p_x\phi_\bfp|^2\Big) + \sum_{\{\mcn_\bfp\}} \cos\(\int \frac{d^{d-1}p}{(\twp)^{d-1}} \mcn_\bfp\phi_\bfp\)\Bigg) + S_{0,\vp}\eea 
		where $\vp$ is the phase of the condensed dipole field, $S_{0,\vp}$ is the free spin-wave action for $\vp$, and where the integrals over all components of $\bfp$ run from $-\pi/a$ to $\pi/a$, with $a$ the lattice spacing. 
		
		The functions $\mcj,\mck,\mcm,\mcn$ appearing in $S$ are required to respect the symmetries of the square lattice, and to be compatible with charge and dipole conservation. Dipole conservation imposes that $\mcm_\bfzero = 0$, but hopping in directions normal to $\uvx$ nevertheless allows for the single derivative terms  $|\p_x\phi_{\bfp}|^2$ to be present for all $\bfp\neq\bfzero$. Conservation of charge, dipole moment, and compactness of $\phi$ requires that $\mcn_\bfp$ be the Fourier transform of an integer-valued function such that $\mcn_\bfzero = 0$ and $(\frac{\p \mcn}{\p p^a})|_{\bfp=\bfzero}=0$. In real space, the simplest terms appearing in the $\mcm_\bfp$ term are $(\p_x \De_b\phi)^2$ with $\De_b$ the discrete lattice derivative along $\uvb\neq\uvx$, while the simplest terms appearing in the cosine are $\De_a\De_b\phi$, with $\uva,\uvb\neq\uvx$.\footnote{If either of $\uva,\uvb$ are equal to $\uvx$, they can be replaced with $a\p_x$, and the cosine can then be Taylor expanded---the resulting term then simply makes a contribution to the $\mcm_\bfp$ term.}
		
		As in the analysis of case where all of the $d^a$ condense, the dipole phase field $\vp$ can effectively be dropped, since after shifting $\vp_\bfp$ by $\p_x\phi_\bfp$ we generate a mass term for $\vp_\bfp$, along with unimportant terms that either are irrelevant, or can be absorbed by a redefinition of $\mcm_\bfp$. Performing this shift, we then write the free term for $\phi_\bfp$ as 
		\be S_{0,\phi} =\int \frac{d^{d-1}p}{(\twp)^{d-1}}  \, \frac{R^2_\bfp}{4\pi} \int d\tau \, dx\, \( \frac1{v_\bfp} |\p_\tau\phi_\bfp|^2 + v_\bfp |\p_x\phi_\bfp|^2\),\ee 
		where we have defined 
		\be R^2_\bfp \equiv \twp \sqrt{\mcj_\bfp \mcm_\bfp},\qq v_\bfp \equiv \sqrt{\mcm_\bfp/\mcj_\bfp}.\ee

		The IR theory described by the above free action contains fields which disperse in a quasi 1d fashion. This quasi 1d behavior will persist so long as cosines containing discrete derivatives along directions normal to $\uvx$ are irrelevant. The simplest of these cosines are $\cos(\De_a^2\phi)$ with $\uva\neq \uvx$ in $d=2,3$ and $\cos(\De_y\De_z\phi)$ in $d=3$, which have scaling dimensions determined by 
		\be \label{sdim} \De_{\cos(\De_a\De_b\phi)} = 8\int \frac{d^{d-1}p}{(\twp)^{d-1}} \frac{\sin^2(p^a/2)\sin^2(p^b/2)}{R^2_\bfp}. \ee 
		Note that while dipole conservation imposes $\mcm_\bfzero = 0 \implies R^2_\bfzero = 0$, the small-$p$ behavior of the numerator means that the integral is still finite. However, the integral diverges in the absense of the sines in the numerator, meaning that the scaling dimension of e.g. $\cos(\phi)$ is infinite (as is the case for all operators which do not conserve dipole moment). 
		
		$R^2_\bfp$ generically increases as we proceed deeper into the single-boson condensed phase, and hence the above terms will eventually become relevant (which happens when their scaling dimensions drops below 2, for the same reasons as explained in \cite{lake2021rg}). 
		When this happens we may replace the discrete derivatives with continuum ones and Taylor expand the cosines, so that the quasi 1d theory (with $z=1$) crosses over to the $z=2$ QLM theory of the BEI discussed in the main text. 
		
		The nature of the transitions where single bosons condense depend as usual on whether or not the transition occurs at fixed density. For the special case where the transition occurs at fixed density, the transition can presumably be identified by determining when the smallest-dimension cosine involving $\t_\bfp$, the field dual to $\phi_\bfp$, becomes irrelevant. Since the bosons are at fixed integer filling, the simplest such translation-invariant cosine is simply $\cos(\t)$, which has scaling dimension 
		\be \De_{\cos(\t)} = \frac12 \int \frac{d^{d-1}p}{(\twp)^{d-1}} R^2_\bfp.\ee 
		Due to the quasi 1d nature of the problem it seems reasonable to expect that in this case, the transition where $\cos(\t)$ becomes irrelevant is of BKT character.
		
		For the generic case of 
		variable density, the transitions are described by a coupled array of dilute Bose gasses, with Lagrangian 
		\bea \mcl &= \sum_\bfl \(\psi^\da_\bfl \p_\tau\psi_\bfl  + \frac1{2m}|(i\D_x-\vp)\psi_\bfl|^2 - \mu |\psi_\bfl|^2\) + \sum_{\bfl,\bfl'} \mcv_{\bfl,\bfl'} |\psi_\bfl|^2 |\psi_{\bfl'}|^2 \\ & \qq + \sum_{\bfl} \( \sum_{a\neq x} (\mcp_{\bfl,a} \psi_{\bfl-\uva}^\da \psi_\bfl^2\psi^\da_{\bfl+\uva} + \sum_{b\neq a \neq x} \mcq_{\bfl,a,b} \psi_{\bfl}^\da \psi_{\bfl+\uva} \psi_{\bfl+\uva+\uvb}^\da \psi_{\bfl+\uvb} + h.c.\) + \mcl_{0,\vp}, \eea  
		where again $\bfl$ indexes coordinates transverse to $\uvx$. The coupling to $\vp$ appears to complicate the analysis of the fixed point slightly, and we defer a detailed RG analysis to future work.

		\section{Effective dipole action} \label{app:effective_action}
		
		In this appendix we derive an effective action for the dipole fields $D^a$ which is valid in the Mott insulating phases, and which allows us to map out at a mean-field level the transitions from the Mott insulators into the dipole condensed phases. 
		
		We begin by writing the hopping term in the microscopic DBHM Hamiltonian \eqref{ham} as 
		\be H_{hop} = -\sum_{i,j,a} b^\da_{i}b_{i+a} [\mca^a]_{ij} b_{j+a}^\da b_j,\ee 
		where the matrix $\mca$ is defined as
		\be [\mca^a]_{ij} = \sum_b \(t\d_{a,b} + t' (1-\d_{a,b}) \)(\d_{i,j+b} + \d_{i,j-b}).\ee 
		As in the main text, we may then decouple the hopping term in terms of dipole fields $D_i^a$ as 
		\be H_{hop} = -\sum_{i,a} \( b_i^\da b_{i+a} D^a_i + (D^a_i)^\da b^\da_{i+a} b_i\) + \sum_{i,j,a} (D^a_i)^\da [\mca^a]\inv_{ij} D_j^a.\ee 
		
		We are now interested in integrating out the boson fields $b_i$ to produce an effective IR action for the dipole fields. On general grounds we may write the dipole action as 
		\bea \label{app_dact} S & = \int d\tau \(\frac w2 \sum_a  |\p_\tau D^a|^2 + \frac12 \sum_{a,b} K_D^{ab} |\D_a D^b|^2 + r \sum_a |D^a|^2 + \frac12 \sum_{a,b} u_{ab} |D^a|^2 |D^b|^2\) + \cdots \\ 
		& \equiv S_2 + S_4 + \cdots,\eea  
		where $S_n$ denotes the terms containing $n$ powers of the dipole fields. These are given explicitly by 
		\bea S_2 & = - C_2 + \int d\tau \sum_{i,j,a} (D^a_i)^\da [\mca^a]_{ij}\inv D^a_j  \\ 
		S_4 & = -C_4 + \frac12 C_2^2,\eea 
		where 
		\bea C_n & \equiv \frac1{n!}\int\prod_{i=1}^n d\tau_i\, \lan \mct [ \prod_{j=1}^n H_{Db}(\tau_j)]\ran\eea 
		with $H_{Db} \equiv - \sum_{i,a} b_i^\da b_{i+a} D^a_i + h.c$, and where the expectation value above is taken with respect to the ground state of the site-diagonal Mott insulating Hamiltonian $H_{onsite} = \sum_i (-\mu n_i + Un_i(n_i-1)/2)$. In what follows we will assume $\mu$ is chosen so that the ground state of $H_{onsite}$ is a Mott insulator with $n>0$ bosons per site. 
		
		We first calculate $C_2$ as 
		\bea C_2 & = \int\frac{d\o}\twp \sum_{i,a} |D_i^a(\o)|^2\int d\tau\, e^{i\o \tau}\, \lan T[(b^\da_i b_{i+a})(\tau) (b_i b^\da_{i+a})(0)]\ran \\ 
		& = \int\frac{d\o}\twp \sum_{i,a} |D_i^a(\o)|^2\int d\tau\, e^{i\o \tau}\sum_l \( \ct(\tau) e^{-\tau (E_l -E_0)} |\lan 0| b^\da_i b_{i+a} |l\ran |^2 + \ct(-\tau)  e^{\tau (E_l - E_0)} |\lan 0|b_{i+a}^\da b_i |l\ran |^2 \) \\ 
		& = \int\frac{d\o}\twp \sum_{i,a} |D_i^a(\o)|^2\sum_l |\lan 0|b^\da_i b_{i+1} |l\ran|^2 \( \frac1{i\o + E_l - E_0} + \frac1{-i\o + E_l - E_0} \),
		\eea 
		where $E_0$ is the ground state energy of $H_{onsite}$ and $l$ runs over all of $H_{onsite}$'s eigenstates. Since the energy of a particle hole excitation above the Mott insulator is always $U$ regardless of $n$ or $\mu$, we may expand in small $\o \ll U$ and write 
		\bea C_2 = \frac{2n(n+1)}U\int \frac{d\o}\twp \sum_{i,a} |D_i^a(\o)|^2 \( 1 - \frac{\o^2}U\).\eea 
		This determines the coefficient $w$ of the time derivative term appearing in \eqref{app_dact} as 
		\be w = \frac{4n(n+1)}{U^2}.\ee 
		Note that as claimed, no linear time derivative term of the form $(D^a)^\da \p_\tau D^a$ appears, due to the particle-hole symmetry present in the expression for $C_2$.   
		
		To derive $r$ and $K_D^{ab}$, we use the fact that 
		\be \sum_{j,m} [\mca^a]\inv_{j,m} e^{i\bfj \cdot \bfq + i\bfm\cdot\bfp} = \frac{\d_{\bfq,-\bfp}}{2\sum_b \(t\d_{a,b} + t' (1-\d_{a,b})\) \cos(p_b)} \ee 
		to obtain 
		\be r = \frac1{2 \(dt + (d-1)t'\)} - \frac{2n(n+1)}U\ee 
		and 
		\be K_D^{ab} = \frac1{dt + (d-1)t'}\( t\d^{ab} + \frac{t'}2(1-\d^{ab})\).\ee 
		
		Now for $C_4$. For the purposes of determining the most relevant terms in $S_4$, we may select out the part of $C_4$ which is local in time. This is the part that provides the fourth-order correction to the ground state energy of $H_{onsite}$ when perturbing in powers of $H_{Db}$, and a straightforward calculation gives 
		\bea \label{cfour} S_4 & = \int d\tau \, \( \sum_{lmn\neq0} \frac{\mco_{0l}\mco_{lm}\mco_{mn}\mco_{n0}}{E_{0l}E_{0m}E_{0n}} - \sum_{lm\neq0} \frac{|\mco_{0l}|^2 |\mco_{0m}|^2 }{E_{0l}^2 E_{0m}} \) \eea 
		where $E_{0p} \equiv E_0-E_p$ and where we have defined the operator 
		\be \mco \equiv \sum_{i,a} (b^\da_ib_{i+a})(0) D^a(\tau) + h.c.\ee 
		
		The evaluation of $S_4$ by way of \eqref{cfour}, and hence the determination of $u_{ab}$, is straightforward but tedious, and here we only quote the result. Dropping derivatives of $D^a$ and writing $u_{ab} = \d_{ab} u_d + (1-\d_{ab})u_o$, we find 
		\bea u_d & = \frac{n(n+1)(4+n(n+1))}{3U^3} \\ 
		u_o & = -\frac{4n(n+1)(10+19n(n+1))}{3U^3}. \eea 
		Note that $u_o<0$, so that in mean field the system favors condensation of all components of $D^a$, and that $|u_o|>u_d$, so that the potential as derived in mean-field is unbounded from below, most likely leading to a first-order transition (this conclusion is of course non-universal, however). 
		
		\bs 
		
	\end{widetext}

\end{document}